\documentclass{emulateapj}
\usepackage{apjfonts}
\usepackage{natbib}

\newcommand{\chandra}{{\it Chandra\/}}
\newcommand{\flux}{{erg~cm$^{-2}$~s$^{-1}$}}
\newcommand{\lum}{{erg~s$^{-1}$}}
\newcommand{\mlum}{{erg~s$^{-1}$~Hz$^{-1}$}}

\begin{document}

\title{Revealing a Population of Heavily Obscured Active Galactic Nuclei
at $z\approx0.5\textrm{--}1$ in the {\bf {\em Chandra}} Deep Field-South}

\author{
B.~Luo,\altaffilmark{1,2,3}
W.~N.~Brandt,\altaffilmark{1,2}
Y.~Q.~Xue,\altaffilmark{1,2}
D.~M.~Alexander,\altaffilmark{4}
M.~Brusa,\altaffilmark{5}
F.~E.~Bauer,\altaffilmark{6}
A.~Comastri,\altaffilmark{7}
A.~C.~Fabian,\altaffilmark{8}
R.~Gilli,\altaffilmark{7}
B.~D.~Lehmer,\altaffilmark{9,10}
D.~A.~Rafferty,\altaffilmark{11}
D.~P.~Schneider,\altaffilmark{1}
\& C.~Vignali\altaffilmark{12}}
\altaffiltext{1}{Department of Astronomy \& Astrophysics, 525 Davey Lab,
The Pennsylvania State University, University Park, PA 16802, USA}
\altaffiltext{2}{Institute for Gravitation and the Cosmos, The Pennsylvania State 
University, University Park, PA 16802, USA}
\altaffiltext{3}{Harvard-Smithsonian Center for Astrophysics, 60 Garden Street, 
Cambridge, MA 02138, USA}
\altaffiltext{4}{Department of Physics, Durham University,
Durham, DH1 3LE, UK}
\altaffiltext{5}{Max-Planck-Institut f\"ur Extraterrestrische Physik,
Giessenbachstrasse, D-85748 Garching b. M\"unchen, Germany}
\altaffiltext{6}{Pontificia Universidad Cat\'{o}lica de Chile,
Departamento de Astronom\'{\i}a y Astrof\'{\i}sica, Casilla 306,
Santiago 22, Chile}
\altaffiltext{7}{INAF---Osservatorio Astronomico di Bologna, Via Ranzani 1,
Bologna, Italy}
\altaffiltext{8}{Institute of Astronomy, Madingley Road, Cambridge,
CB3 0HA, UK}
\altaffiltext{9}{The Johns Hopkins University, Homewood Campus, Baltimore, MD
21218, USA}
\altaffiltext{10}{NASA Goddard Space Flight Centre, Code 662, Greenbelt, MD 20771, USA}
\altaffiltext{11}{Leiden Observatory, Leiden University, Oort Gebouw, P.O. Box
9513 RA, Leiden, The Netherlands}
\altaffiltext{11}{Universit\'a di Bologna, Via Ranzani 1, Bologna, Italy}

\begin{abstract}
Heavily obscured ($N_{\rm H}\ga3\times10^{23}$~cm$^{-2}$) Active Galactic Nuclei
(AGNs) not detected
in even the deepest X-ray surveys are often considered to be comparably numerous to 
the unobscured and moderately obscured AGNs. Such sources are required to
fit the cosmic X-ray background (XRB) emission in the 
10--30 keV band. We identify 
a numerically significant population of heavily obscured AGNs 
at $z\approx0.5$--1 in the \chandra\ Deep \hbox{Field-South} (CDF-S)
and Extended \chandra\ Deep Field-South
by selecting 242 X-ray undetected objects with
infrared-based star formation 
rates (SFRs) substantially higher (a factor of 3.2 or more) 
than their SFRs determined from the UV after correcting for dust extinction. 
An \hbox{X-ray}
stacking analysis of 23 candidates in the central CDF-S region using the 4 Ms 
\chandra\ data 
reveals a hard X-ray signal with an effective power-law photon
index of $\Gamma=0.6_{-0.4}^{+0.3}$, indicating
a significant contribution from obscured AGNs. Based on Monte Carlo simulations,
we conclude that 
$74\pm25\%$ of the selected galaxies host obscured
AGNs, within which $\approx95\%$ are heavily obscured and $\approx80\%$ 
are Compton-thick (CT; $N_{\rm H}>1.5\times10^{24}$~cm$^{-2}$). 
The heavily obscured objects in our sample are of moderate intrinsic 
X-ray luminosity
[$\approx(0.9$--$4)\times10^{42}$~\lum\ in the 2--10 keV band].
The space density of the CT AGNs 
is $(1.6\pm0.5)\times10^{-4}$~Mpc$^{-3}$.
The $z\approx0.5$--1 CT objects studied here are expected to contribute $\approx1\%$
of the total XRB
flux in the 10--30 keV band, and they account for
$\approx5$--15\% of the emission in this energy band expected from all CT AGNs
according to population-synthesis models.
In the 6--8 keV band, the stacked signal of the 23 heavily obscured 
candidates accounts for $<5\%$ of the unresolved XRB flux, while 
the unresolved $\approx25\%$ of the XRB 
in this band can probably be explained by a stacking
analysis of the 
X-ray undetected optical galaxies in the CDF-S (a 2.5~$\sigma$ stacked signal).
We discuss prospects to identify such heavily obscured objects
using future hard X-ray
observatories.
\end{abstract}

\section{INTRODUCTION}
Deep X-ray surveys have provided the most effective method of identifying
reliable and fairly complete 
samples of Active Galactic Nuclei (AGNs) out to $z\approx5$ (e.g., see 
\citealt{Brandt2005} for a review). The observed AGN sky density
reaches $\approx10\,000$~deg$^{-2}$ in the deepest X-ray surveys, the 
\chandra\ Deep Fields \citep[e.g.,][]{Bauer2004,Xue2011}.
These \hbox{X-ray} point sources are largely responsible for the 
observed cosmic X-ray background (XRB). A significant portion 
($\approx70\textrm{--}90\%$) of the XRB in the \hbox{0.5--6} keV range
has been resolved into discrete sources by 
\chandra\ and {\it XMM-Newton} \citep[e.g.,][]{Moretti2002,Bauer2004,Hickox2006},
the majority of which are AGNs with moderate X-ray obscuration 
($N_{\rm H}\la3\times10^{23}$~cm$^{-2}$; 
e.g., \citealt{Szokoly2004,Barger2005,Tozzi2006}). 
However, the resolved fraction of 
the XRB decreases toward higher energies, being $\approx60\%$ in the
6--8 keV band and $\approx50\%$ in the 8--12~keV band
\citep[][]{Worsley2004,Worsley2005}. Population-synthesis models
suggest that at $\approx20$--30~keV, where
the XRB reaches its peak flux \citep[e.g.,][]{Gruber1999,Moretti2009}, the unobscured or moderately obscured 
AGNs discovered at lower energies cannot account for the XRB flux entirely,
and an 
additional population of heavily obscured ($N_{\rm H}\ga3\times10^{23}$~cm$^{-2}$), or even Compton-thick
($N_{\rm H}>1.5\times10^{24}$~cm$^{-2}$, hereafter CT), AGNs at mainly 
$z\approx0$--1.5 is 
required. The number density 
of these heavily obscured AGNs 
is estimated to be of the
same order as that of moderately obscured AGNs \citep[e.g.,][]{Gilli2007}. 
However, as the X-ray emission (below at least 10 keV) of such sources is
significantly suppressed,
only a few distant heavily obscured
AGNs have been clearly identified even in the
deepest X-ray surveys
\citep[e.g.,][]{Tozzi2006,Alexander2008,Georgantopoulos2009,Comastri2011,Feruglio2011},
and thus a significant fraction of the AGN population 
probably remains undetected.
 
Infrared (IR) selection is a powerful tool for detecting
heavily obscured AGNs that cannot be identified in the X-ray band. 
X-ray and UV emission absorbed by the obscuring material
is reprocessed and reemitted mainly at mid-IR wavelengths 
\citep[e.g.,][]{Silva2004,Prieto2010}. The mid-IR emission is 
less affected by dust
extinction than at optical or near-IR wavelengths, 
rendering it more suitable for 
identifying obscured AGNs. 
With the {\it Spitzer} mission \citep{Werner2004},
deep mid-IR data in multiple bands can be obtained by 
the Infrared Array Camera (IRAC)
and Multiband Imaging Photometer (MIPS) instruments.

Various IR selection methods for obscured AGNs have been proposed.
{\it Spitzer} IRAC power-law
selection chooses sources whose IRAC spectral energy distributions (SEDs) 
follow a power law with a slope of $\alpha\le-0.5$ 
($f_{\nu}\propto\nu^{\alpha}$; e.g., \citealt{Alonso2006,Donley2007,Donley2010}).
This criterion was determined based on 
the average optical-to-IR spectral properties
of X-ray selected QSOs \citep[e.g.,][]{Elvis1994}. 
It can select a relatively pure sample of AGNs with little contamination
from star-forming galaxies but is 
generally limited to the most-luminous sources \citep[e.g.,][]{Polletta2008}. 
IRAC color-color selection
applies color cuts in the IRAC color space to identify AGNs 
\citep[e.g.,][]{Lacy2004,Stern2005}. This method 
is based on the IRAC color properties of optically selected QSOs and AGNs, 
and it tends to have 
significant contamination from star-forming galaxies when applied to deep
mid-IR data \citep[e.g.,][]{Cardamone2008,Georgantopoulos2008,Brusa2010}. 
IR-excess selection uses a combined UV--optical--IR 
color cut to identify AGNs \citep[e.g.,][]{Dey2008,Fiore2008,Lanzuisi2009}. 
This technique is also contaminated by star-forming galaxies
\citep[e.g.,][]{Donley2008}.

Compared to the above methods, 
which utilize observational data directly and determine the selection
criteria empirically,
\citet{Daddi2007}
adopted a generally different approach to find obscured AGNs. They selected 
candidates with significantly higher 
IR-based (also including the component corresponding to 
the transmitted UV emission; see \S2.1 below for details) 
star formation rates (SFRs) than their UV-based 
(after correcting for dust extinction) SFRs.
The amount of excess in the IR-based SFR is a measurement of  
the excess in the IR emission, which is likely caused by
reprocessed emission from the dusty torus 
of a heavily obscured AGN.
This relative IR SFR excess (ISX) selection method
requires and utilizes redshift information to measure
the IR excess effectively, and it 
appears to
isolate a different IR sample
from the methods above \citep{Donley2008,Alexander2008}. 
A few of the candidates in \citet{Daddi2007} have been reliably identified as 
CT AGNs \citep[][]{Alexander2008}. 
The ISX sample is also contaminated by star-forming galaxies
due to uncertainties in the calculation of SFRs; the 
heavily obscured AGN fraction in the \citet{Daddi2007}
sample is estimated to be $\approx25\%$ \citep{Alexander2011}, 
and the rest of the objects are likely star-forming galaxies.

In this paper, we improve the ISX selection method and utilize it to
search for \hbox{X-ray} undetected, heavily obscured AGNs at $z\approx0.5$--1
in the \chandra\ Deep Field-South
(\hbox{CDF-S}; \citealt{Giacconi2002,Luo2008,Xue2011})
and Extended \chandra\ Deep \hbox{Field-South}
(E-CDF-S; \citealt{Lehmer2005}). We choose to apply the ISX method on
the CDF-S and E-CDF-S galaxy sample in this redshift range
for the following reasons:
\begin{itemize}
\item {\it IR-selection of heavily obscured AGNs at $z\approx0.5\textrm{--}1$ is a
poorly explored territory.} Most of the previous IR selections were focused on
samples at $z\approx2$, including the \citet{Daddi2007} work. 
Since AGNs at $z\approx0.5$--1
contribute significantly to the resolved XRB \citep[e.g.,][]{Worsley2005,Gilli2007},
and heavily obscured AGNs in this redshift range are expected to play a similar
role in the missing fraction of the XRB \citep[e.g.,][]{Gilli2007,Treister2009},
it is of interest to study the heavily obscured population at these lower
redshifts. 
\citet{Fiore2009} selected a sample of 
IR-excess sources in the COSMOS field, which included some objects at $z<1$. 
However, detailed studies (e.g., the AGN fraction) were not focused on this
low-redshift bin.
Moreover, the X-ray and IR (24~$\mu$m) flux limits in the CDF-S field are 
$\ga10$ times more sensitive than those in the COSMOS field \citep{Fiore2009},
allowing identification of the population of 
heavily obscured AGNs with lower intrinsic luminosities (which
are likely more numerous).
Recently, \citet{Georgakakis2010}
selected a sample of 19 IR-excess sources at $z\approx1$ 
in the AEGIS and GOODS-N fields, but
they argued that most of those X-ray undetected candidates are
not heavily obscured AGNs based on IR SED modeling.

\item
{\it The ISX method is best applicable at $z\approx0.5\textrm{--}1$.} 
The ISX method proposed by \citet{Daddi2007} is a promising technique
because of its simple physical motivation and its success in
selecting identified CT AGNs at $z\approx2$. 
However, the ISX method has not been applied to any other 
studies, partially due to its known limitations. There are two essential
quantities in the ISX method, the IR-based SFR ($\textrm{SFR}_{\rm IR+UV}$) and
the UV-based SFR ($\textrm{SFR}_{\rm UV,corr}$). The IR-based SFR is derived 
mainly from the observed {\it Spitzer} MIPS 24~$\mu$m flux
using the \citet{Chary2001} galaxy SED templates. It has been
noted that at $z>1.4$, the IR luminosity (and thus $\textrm{SFR}_{\rm IR+UV}$) derived
this way
is overestimated by an average factor of $\approx5$
\citep{Murphy2009,Murphy2011},\footnote{This is
largely due to the fact that the local high-luminosity SED templates
in \citet{Chary2001} do not
accurately account for
the aromatic features around 24~$\mu$m and the IR SEDs at $z>1.4$;
they show weaker PAH emission than that in high-redshift galaxies by
an average factor of $\approx5$.} leading to
additional contamination from galaxies in the ISX sample.
The UV-based SFR is calculated using the dust-extinction corrected UV luminosity.
\citet{Daddi2007} used an empirical color-extinction relation to estimate
the dust extinction. However, this correlation cannot be applied to relatively
old galaxies that are intrinsically red (dominated by old stellar populations
instead of being reddened), and all
these galaxies were removed from the \citet{Daddi2007} sample.\footnote{
\citet{Daddi2007} removed galaxies with SSFR$<$median(SSFR)/3, where 
SSFR is the specific SFR, defined as 
$\textrm{SSFR}=\textrm{SFR}_{\rm IR+UV}/M_{*}$ with $M_{*}$ being the stellar mass.
About 15\% of the sources were excluded this way, the median stellar mass of 
which is $\approx5$ times larger than that of the remaining sources.}
As AGNs tend to be hosted by massive, red galaxies \citep[e.g.,][]{Xue2010},
this limitation might significantly affect the completeness of the
resulting ISX sample. Both these problems
can be substantially alleviated if the ISX method is applied at lower 
redshifts.\footnote{We note that applying the ISX
method at $z\approx2$ has the advantage of being able to get
a better contrast
between star-formation and AGN 
emission at the observed 24~$\mu$m wavelength
than at lower redshifts.} 
The 24~$\mu$m deduced IR luminosity is robust at $z\la1$ \citep[e.g.,][]{Elbaz2010}, while 
the dust extinction can be derived via SED fitting \citep[e.g.,][]{Brammer2009,Reddy2010}.
The \hbox{SED-fitting} technique is reliable with the high-quality multiwavelength photometric data achievable
at $z\la1$.

\item
{\it The CDF-S and E-CDF-S are excellent fields 
in which to perform such a study.} 
They have been covered by extensive multiwavelength photometric 
and spectroscopic surveys, including very deep X-ray and 24~$\mu$m exposures.
The COMBO-17 \citep{Wolf2004,Wolf2008} and Subaru \citep{Cardamone2010} surveys
covering the entire CDF-S and E-CDF-S
are particularly valuable for the determination of dust extinction via SED fitting.
The 17-filter coverage of COMBO-17 between $3600$~\AA\ and $9200$~\AA\ 
and 18 medium-band coverage of 
Subaru between $4200$~\AA\ and $8600$~\AA\ are useful
for distinguishing between a red dust-free galaxy and a blue dusty galaxy.
Reliable photometric redshifts (photo-$z$'s) can also be obtained with the 
high-quality multiwavelength data available.

\item
{\it We can assess the possibility of directly detecting the missing population of
heavily obscured AGNs with
future hard X-ray observatories.} 
One of the science goals of several future hard X-ray missions, 
such as the {\it Nuclear
Spectroscopic Telescope Array} ({\it NuSTAR}; \citealt{Harrison2010}) 
and {\it ASTRO-H} \citep{Takahashi2010}, is
to resolve better the XRB at $\approx10$--30~keV via
detecting heavily obscured and CT AGNs directly in the distant universe. While IR-selected
heavily obscured AGN candidates at $z\approx2$
are likely below their sensitivity limits,\footnote{Assuming an absorption 
column density
of $N_{\rm H}=10^{24}$~cm$^{-2}$,
the 10--30 keV flux of the intrinsically luminous (\hbox{2--10} keV 
luminosity of $2\times10^{44}$~\lum) CT AGN HDF-oMD49 
at $z=2.21$ \citep{Alexander2008} will still be below the 
detection limit of {\it NuSTAR} with a 1~Ms exposure (the {\it NuSTAR}
sensitivity limit is from \url{http://www.nustar.caltech.edu/}). The 
average luminosity of the CT AGN candidates in \citet{Daddi2007} is about 
an order of magnitude fainter than that of HDF-oMD49.} candidates at $z\approx0.5$--1 
should be more easily detectable, and we will critically assess such possibilities.

\end{itemize}

This paper is organized as follows.
We describe
the multiwavelength data and ISX sample selection in \S2.
In \S3 we present an X-ray stacking analysis of the ISX sample 
and estimate the heavily obscured AGN fraction among the sample via simulations. In \S4,
we calculate the space density of the selected heavily obscured AGNs and 
their contribution to the XRB.
We also discuss the feasibility
of detecting these ISX sources with future hard X-ray
observatories. We
summarize in \S5.
Throughout this paper, 
we adopt the latest cosmology with
$H_0=70.4$~km~s$^{-1}$~Mpc$^{-1}$, $\Omega_{\rm M}=0.272$,
and $\Omega_{\Lambda}=0.728$ \citep{Komatsu2010}. All given
magnitudes are in the AB system \citep[e.g.,][]{Oke1983} unless otherwise stated.

\section{THE RELATIVE IR SFR EXCESS SAMPLE}
\subsection{Multiwavelength Data and Source Properties}
The CDF-S is the deepest X-ray survey ever performed, having 
a total \chandra\ exposure of $\approx4$~Ms and
covering a solid angle of $\approx460$~arcmin$^2$ \citep{Xue2011}.
X-ray images in three standard bands, \hbox{0.5--8.0~keV} (full band; FB),
\hbox{0.5--2.0~keV}
(soft band; SB), and \hbox{2--8~keV} (hard band; HB), along with 
other relevant products such as the main X-ray
source catalog (740 sources) and sensitivity maps,
are used in the sample creation and X-ray stacking analysis, which are discussed below 
and in \S3. 
The CDF-S is flanked by the E-CDF-S, which consists of four contiguous
$\approx250$~ks \chandra\ observations with
a total solid angle of $\approx1100$~arcmin$^2$ \citep{Lehmer2005}.
The E-CDF-S main X-ray source catalog (762 sources) 
is used only in the sample creation.

Mid-IR-to-optical multiwavelength data are required to 
calculate the IR-based and \hbox{UV-based}
SFRs in the ISX method. The CDF-S and E-CDF-S have been covered 
by extensive photometric and spectroscopic
surveys, and we constructed a sample of mid-IR and optically
selected sources in the CDF-S and E-CDF-S region.
For the mid-IR data, we used the
{\it Spitzer} MIPS 24~$\mu$m source catalog
from the {\it Spitzer} Far Infrared Deep Extragalactic
Legacy Survey (FIDEL; M. Dickinson et~al. 2011, in preparation).
The FIDEL survey covers the entire CDF-S and \hbox{E-CDF-S}, and it 
has a 5~$\sigma$ limiting flux of
$\approx20$~$\mu$Jy. 
For the \hbox{IR-to-optical} data,
we used the master source catalog compiled by \citet{Rafferty2010}, which
consists of $\approx100\,000$ optically selected galaxies covering 
the entire CDF-S and \hbox{E-CDF-S}.
The master source catalog 
was created based on the MUSYC catalog \citep{Gawiser2006},
the COMBO-17 catalog \citep{Wolf2004,Wolf2008}, and the GOODS-S MUSIC catalog 
\citep{Grazian2006}, and it was also cross-matched to several other photometric 
catalogs such as the 
SIMPLE IRAC catalog \citep{Damen2010} and the {\it GALEX} UV catalog
\citep[e.g.,][]{Morrissey2005} 
to include up to 42 bands of IR-to-UV data.
Additionally, we included the 18 medium-band Subaru photometric data
that have become available recently \citep{Cardamone2010};
about 80\% of the sources in the master catalog have Subaru data. 
For our purpose of computing reliably
the UV-based SFRs (requiring robust \hbox{IR-to-UV} SED fitting),
we chose sources with relatively bright $R$-band
magnitudes ($R<25$); there are $\approx40\,000$ such objects. 
These sources were matched to the 24~$\mu$m sources using the 
likelihood-ratio matching technique described in \citet{Luo2010}, 
resulting in 5237 matches 
and a false-match probability of $\approx4\%$.
For
sources in the X-ray stacking samples discussed in \S3, we
have visually examined their IR and optical images and removed sources
that are probably affected by source blending, and 
thus the false-match probability
is negligible for those samples.

Reliable spectroscopic redshifts (spec-$z$'s; 979/5237 sources) 
for sources in the CDF-S and E-CDF-S 
were collected 
from the following
catalogs: \citet{LeFevre2004},
\citet{Szokoly2004},
\citet{Mignoli2005},
\citet{Ravikumar2007}, \citet{Vanzella2008}, \citet{Popesso2009},
\citet{Balestra2010}, and \citet{Silverman2010}. 
If a spec-$z$ is not available for a given source, 
we calculated its photo-$z$ using 
the Zurich Extragalactic Bayesian Redshift Analyzer (ZEBRA;
\citealt{Feldmann2006}).
ZEBRA utilizes a maximum-likelihood
approach to find the best-fit SED template; 
more details about the ZEBRA SED-fitting procedure are discussed in
\S3.2 of \citet{Luo2010}.
For the purpose of our study here, we used ZEBRA to derive photo-$z$'s
and dust extinction (in terms of the extinction in the $V$
band, $A_V$) simultaneously using up to 60 photometric bands. 
We selected 99 typical
PEGASE galaxy templates from \citet{Grazian2006}
that span a wide range of star-formation history, and we
applied intrinsic extinction in
the range $A_V=0$--4 with an increment of 0.1 to these templates 
employing the \citet{Calzetti2000} extinction law.
The resulting
SED templates were used to fit the \hbox{IR-to-UV} SED data (excluding the 24 $\mu$m data)
of the sources.
For sources with spec-$z$'s, 
the redshifts were fixed at the spec-$z$ values during the SED fitting, and
the values of $A_V$
were obtained from the best-fit templates.
For the other sources, both the photo-$z$ and $A_V$ values
were determined based on the best-fit templates.
To check the quality of the photo-$z$'s, we performed 
another ZEBRA run of the spec-$z$ sources, setting the redshift
as a free parameter, and then compared the resulting photo-$z$'s with the spec-$z$'s. 
The photo-$z$ accuracy was estimated using the
outlier fraction and 
the normalized median absolute deviation 
($\sigma_{\rm NMAD}$; e.g., \citealt{Brammer2008,Luo2010}) parameters. 
Outliers are defined as sources having 
\hbox{$|\Delta z|/(1+z_{\rm spec})>0.15$}, where
$\Delta z=z_{\rm photo}-z_{\rm spec}$, and $\sigma_{\rm NMAD}$ is defined as
\begin{equation}
\sigma_{\rm NMAD}=1.48\times {\rm median}\left(\frac{|\Delta z-
{\rm median}(\Delta z)|}{1+z_{\rm spec}}
\right)~.
\end{equation}
For all the spec-$z$ sources, 
we found an outlier fraction of 6\% and $\sigma_{\rm NMAD}=0.020$.
For sources in the redshift range of 0.5--1, which are of primary interest for this
study, the outlier fraction is 3\% and $\sigma_{\rm NMAD}=0.017$.
These photo-$z$'s have comparably high quality to those recently obtained for 
galaxies in the CDF-S region \citep[][]{Cardamone2010, Dahlen2010}.

We calculated the IR-based and UV-based SFRs following \citet{Bell2005},
\begin{equation}
\textrm{SFR}_{\rm IR+UV}=9.8\times10^{-11}(L_{\rm IR}/L_{\sun}+L_{\rm UV}/L_{\sun})~~M_{\sun}~{\rm yr}^{-1},
\end{equation}
and
\begin{equation}
\textrm{SFR}_{\rm UV,corr}=9.8\times10^{-11}(L_{\rm UV,corr}/L_{\sun})~~M_{\sun}~{\rm yr}^{-1},
\end{equation}
where $L_{\sun}=3.8\times10^{33}$~erg~s$^{-1}$ and $L_{\rm IR}$, $L_{\rm UV}$, and
$L_{\rm UV,corr}$ are the IR, UV, and \hbox{dust-extinction} corrected UV
luminosities, respectively. 
The \citet{Kroupa2001} initial mass function was adopted here.
The IR luminosity was estimated by 
finding an IR SED template that produces
the observed 
24~$\mu$m luminosity via interpolation of a library of 105 IR SEDs 
and then calculating
the integrated 8--1000 $\mu$m luminosity for this template \citep{Chary2001}.
It has been demonstrated that the IR luminosity determined this way 
is 
similar to that derived with additional longer wavelength
photometric data and mid-IR spectroscopic for sources 
with $z\la1.4$ and $L_{\rm IR}\la3\times10^{12}~L_{\sun}$ \citep{Murphy2009,Murphy2011}.
The UV luminosity was computed
following $L_{\rm UV}=3.3\nu l_{\nu, 2800{\rm \AA}}$ (see 
\S3.2 of \citealt{Bell2005}), where the rest-frame 2800~\AA\
monochromatic luminosity $l_{\nu, 2800{\rm \AA}}$ was interpolated from the 
multiwavelength data. We calculated the \hbox{dust-extinction} correction for the UV luminosity
employing the \citet{Calzetti2000} extinction law, 
$L_{\rm UV,corr}=10^{0.72A_V}L_{\rm UV}$.

As AGNs tend to
reside in massive galaxies \citep[e.g.,][]{Kauffmann2003,Brusa2009,Xue2010},
we estimated stellar masses for the 5237 galaxies and used these masses to 
filter out low-mass objects.
The stellar mass is calculated
following \citet{Xue2010}:
\begin{equation}
\log(M_{*}/M_{\sun})=\log(L_{K}/L_{\sun,K})+b_{K}(M_B-M_V)+a_{K}-0.10~,
\label{smass}
\end{equation}
where $L_{\sun,K}$ is the monochromatic $K$-band luminosity of the Sun ($L_{\sun,K}=3.6\times10^{18}$~\mlum),
the coefficients $a_{K}=-1.390$ and $b_{K}=1.176$ are from
\citet{Zibetti2009}, and the \hbox{rest-frame} monochromatic luminosity $L_{K}$ and color
$(M_B-M_V)$ in the Vega system 
are from the \hbox{SED-fitting} results. All the sources have IRAC detections, and thus
they all have \hbox{rest-frame} $K$-band coverage in the SED fitting.
The normalization has been adjusted by
$-0.10$ dex to account for our adopted \citet{Kroupa2001} initial
mass function. 
The $K$-band luminosity was used because it is 
$\approx5$--10 times less sensitive to dust and stellar-population effects
than optical luminosities \citep[e.g.,][]{Bell2000}.

\subsection{Sample Selection}
We define our parent sample from the 5237
optically and 24~$\mu$m selected galaxies above
with the following criteria: (1)
The redshift of the source is between 0.5 and 1; there are 2037 such sources.
(2) The source must have 
COMBO-17 or Subaru detections (in $>10$ photometric bands) 
to ensure robust SED fitting;
there are 24 sources removed by this criterion.
(3) The source has a 
stellar mass $M_{*}>5\times10^9~M_{\sun}$, as AGNs tend to 
be hosted by massive galaxies \citep[e.g.,][]{Kauffmann2003,Brusa2009,Xue2010};
this criterion removes $\approx25\%$ of the sources.
(4) The source is not X-ray detected or overlapping with the 
90\% encircled-energy
aperture (see \S3.2 of \citealt{Xue2011}) of any known X-ray source;
the latter criterion is to avoid any contamination from nearby X-ray sources
in our X-ray stacking analysis.
We used the 4~Ms CDF-S and 250~ks E-CDF-S X-ray source catalogs
for this purpose \citep{Lehmer2005,Xue2011}. 
About 200 sources are removed by this criterion.
We note that the basic source properties,
such as the dust extinction, stellar mass, and UV-based SFR, were calculated based
on the assumption that
the optical SED is dominated by the host galaxy; this assumption 
is valid after we exclude
the X-ray sources from the sample (e.g., see \S4.6.3 of \citealt{Xue2010}
for further discussion).

\begin{figure}
\centerline{
\includegraphics[scale=0.5]{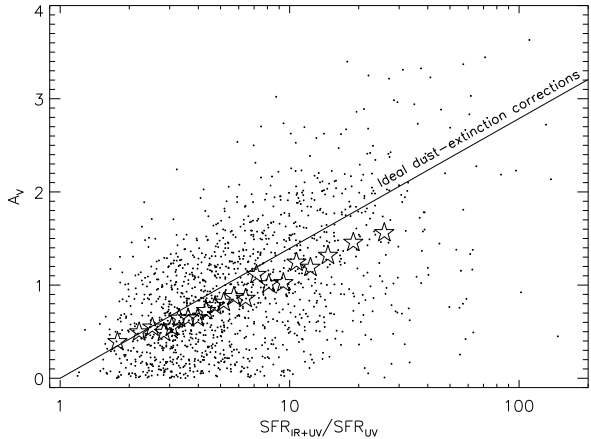}
}
\figcaption{
The dust extinction ($A_V$) derived from SED fitting vs. the ratio of
the IR-based ($\textrm{SFR}_{\rm IR+UV}$; Eq.~2) and UV-uncorrected [$\textrm{SFR}_{\rm UV}=9.8\times10^{-11}(L_{\rm UV}/L_{\sun})~M_{\sun}~{\rm yr}^{-1}$] SFRs for sources in
the parent sample.
The stars indicate the mean $A_V$ values in different $\textrm{SFR}_{\rm IR+UV}/\textrm{SFR}_{\rm UV}$ bins; each bin contains 60 sources. The errors on the mean values are
smaller than or comparable to the symbol size.
The solid line shows the ideal dust-extinction corrections which would
lead to consistent IR-based and UV-based SFRs. The SED-derived $A_V$ values
scatter around this line; the
deviation of the mean $A_V$ values from the ideal line
at large ratios of the two SFRs is likely
due to the
contribution from ISX sources that still have IR excesses after 
dust-extinction corrections (see Fig.~2).
\label{plotav}}
\end{figure}

The parent sample defined above consists of 1313 sources, about 20\% of which
have \hbox{spec-$z$'s}. The spectroscopic completeness does not have any significant 
dependence on the IR or optical magnitude, and it is largely limited by 
the spectroscopic coverage in the area; for example, $\approx60\%$ of the sources 
in the GOODS-S region have spec-$z$'s.  
We show in Figure~\ref{plotav} the dust extinction ($A_V$)
against the ratio of
the IR-based ($\textrm{SFR}_{\rm IR+UV}$) and UV-uncorrected 
[$\textrm{SFR}_{\rm UV}=9.8\times10^{-11}(L_{\rm UV}/L_{\sun})~M_{\sun}~{\rm yr}^{-1}$] SFRs, which is 
also a measure of extinction. The $A_V$ values derived from SED fitting
can well account for the extinction in general. 
In Figure~2, we plot the logarithmic ratios of the
IR-based and UV-based SFRs, defined as
$R_{\rm SFR}=\log(\textrm{SFR}_{\rm IR+UV}/\textrm{SFR}_{\rm UV,corr})$.
The distribution of 
$R_{\rm SFR}$ displays an
excess on the positive side ($R_{\rm SFR}>0$) similar 
to that observed in \citet{Daddi2007} for $z\approx2$ galaxies. 
Various statistical errors are present in the calculations of 
$\textrm{SFR}_{\rm IR+UV}$ and $\textrm{SFR}_{\rm UV,corr}$, such as the 
uncertainties in $L_{\rm IR}$, $L_{\rm UV}$, and $A_V$. 
These tend to make the $R_{\rm SFR}$ distribution 
follow approximately a Gaussian function.
We mirror the negative half of the histogram to the 
positive side, and the combined distribution (blue histogram in Fig. 2) can be approximated by a Gaussian function with $\sigma=0.33$.
We thus consider that the excess in the distribution of $R_{\rm SFR}$ 
is caused by galaxies
hosting heavily obscured or even CT AGNs. We define ISX sources 
using the same criterion as in \citet{Daddi2007}:
$R_{\rm SFR}>0.5$.
For comparison, we define IR SFR normal (ISN) sources as those having
$R_{\rm SFR}<0.2$. There are 
242 ISX and 736 ISN sources in the parent sample.
Assuming that the intrinsic
dispersion of $R_{\rm SFR}$ is Gaussian (blue histogram in Fig. 2), 
and that the excess IR emission is powered by a heavily obscured AGN,
about 73\%
of the ISX sources (176 objects out of the 242 ISX sources; 
shaded region in Fig.~2) host such AGNs.
For simple comparison, we selected X-ray detected sources in the same way from 
the initial 5237 sources
(revising the fourth criterion above to require \hbox{X-ray} detection in the CDF-S
or E-CDF-S source catalogs); 
there are 198 X-ray
sources selected. Following the AGN classification scheme in \S4.4 of
\citet{Xue2011}, 
including criteria for intrinsic X-ray luminosity,
effective power-law photon index, and X-ray-to-optical flux ratio,
we identified 155 AGNs from the X-ray sources, 
mostly unobscured or moderately obscured.
This number is comparable to that
of expected heavily obscured AGNs (176 objects) in the ISX sample, 
consistent with predictions from 
population-synthesis models \citep[e.g.,][]{Gilli2007}.

The ISX and ISN sources have different physical properties. 
The median stellar mass, IR luminosity, and IR-based SFR of 
the ISX sample are 2.2, 1.4, and 1.2 times those for the ISN sample, 
respectively, while the median UV luminosity, UV-based SFR, and $A_V$ of the ISX sample
are 50\%, 15\%, and 30\% those for the ISN sample.
If the relative excess in the IR luminosities is contributed by
underlying heavily obscured AGNs, then such objects
selected by the ISX method at $z\approx0.5$--1 
appear to be hosted by massive galaxies with little dust.
Since AGNs are preferentially hosted by massive galaxies
\citep[e.g.,][]{Kauffmann2003,Brusa2009,Xue2010}, it is 
natural to find ISX sources in such galaxies.
The dust extinction (in terms of $A_V$) applies to the entire galaxy,
and a small value of $A_V$ does not necessarily conflict with a 
high absorbing column density for a heavily obscured AGN in the nucleus
\citep[e.g.,][]{Maiolino2001,Polletta2008}. 
The finding of smaller $A_V$ values in the ISX sources
may be a selection effect.
Heavily obscured AGNs
at \hbox{$z\approx0.5$--1} are numerically dominated by moderate-luminosity
objects (intrinsic X-ray luminosity 
\hbox{$\approx10^{42}$--$10^{43}$~\lum)} according to
population-synthesis models \citep[e.g.,][]{Gilli2007}, and 
sources hosted by galaxies with less dust (less IR emission from the 
host galaxy)
will have more prominent IR excesses and are more likely to be
identified by IR selection methods.

As a test of whether ISX sources are preferential hosts of
obscured AGNs, we also included 
X-ray detected CDF-S AGNs \citep{Xue2011}
in Figure~2. 
There are 74 CDF-S AGNs selected from the initial 5237 sources 
(revising the fourth criterion above to require \hbox{X-ray} detection in the CDF-S source catalog). We further removed 6 luminous AGNs 
with $L_{0.5\textrm{--}8~{\rm keV}} >10^{43.7}$~\lum\ as the optical SEDs
for such sources may be affected by AGN contamination 
\citep[e.g., \S4.6.3 of][]{Xue2010}.
The level of intrinsic absorption was 
estimated by assuming an underlying
X-ray power-law photon index of $\Gamma=1.8$ and using 
XSPEC (Version 12.5.1; \citealt{Arnaud1996}) to derive the appropriate $N_{\rm H}$ 
value
that produces the observed X-ray band ratio (defined as the ratio of count rates
between the HB and SB). We consider a source to be obscured if 
$N_{\rm H}>10^{22}$~cm$^{-2}$; otherwise, it is unobscured or only weakly
obscured.
Figure~2 shows clearly that AGNs in the ISX ($R_{\rm SFR}>0.5$) region are mostly
obscured (17 cases out 19), while only half of the 
AGNs in the ISN region are obscured.
We caution that the nature of the ISX
population as a whole is likely different from these X-ray detected AGNs, given the apparent 
difference in the IR-to-X-ray flux ratios, and thus we perform detailed X-ray
studies of the ISX sample in the following section.

\begin{figure}
\centerline{
\includegraphics[scale=0.5]{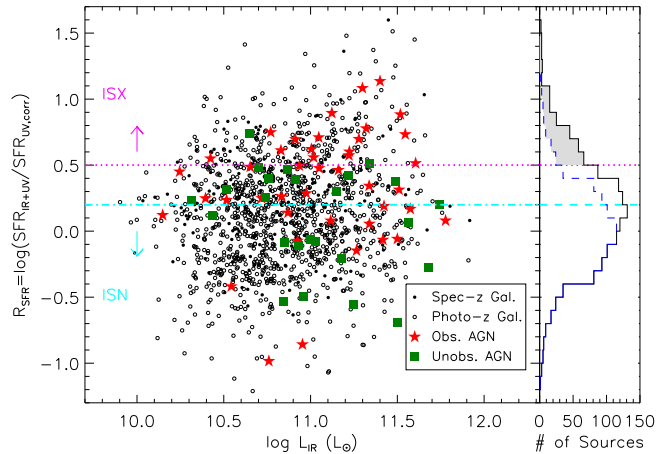}
}
\figcaption{
The logarithmic ratio 
of $\textrm{SFR}_{\rm IR+UV}$ and $\textrm{SFR}_{\rm UV,corr}$  vs. the IR luminosity
for sources in the parent sample. The filled and open dots
represent sources with spec-$z$'s and \hbox{photo-$z$'s}, respectively.
X-ray AGNs in the 4~Ms CDF-S main catalog are shown as
red stars (obscured) and green squares (unobscured or weakly obscured). 
The classification of obscured/unobscured AGNs
is based on X-ray data; see \S2.2.
The right panel shows the distribution of $R_{\rm SFR}$ 
(not counting the X-ray sources). The dashed line is
the reflection of the bottom half of the histogram around $R_{\rm SFR}=0$;
the resulting distribution (the blue histogram) 
can be approximated by a Gaussian function with 
$\sigma=0.33$.
The horizontal dotted and dash-dotted lines indicate our criteria for
defining the ISX and ISN samples, respectively. 
About 73\% of the sources in the ISX
sample (shaded area) 
belong to the excess population compared to the Gaussian distribution, and 
many of these are 
probably hosts of heavily obscured AGNs.
The majority of the X-ray AGNs in the ISX region are obscured.
[{\it See the electronic
edition of the Journal for a color version of this figure.}]
\label{f01}}
\end{figure}

\section{X-RAY STACKED PROPERTIES AND COMPTON-THICK AGN FRACTION}

\subsection{X-ray Stacking Analysis}
X-ray stacking analysis is a useful technique to obtain the average
X-ray properties and probe the nature 
of a sample of X-ray undetected objects, and it has been
used extensively in previous studies of IR-selected AGN candidates 
\citep[e.g.,][]{Daddi2007,Donley2008,Fiore2009}. 
We 
performed X-ray stacking of the ISX and ISN samples in
the three standard X-ray bands of the 4~Ms CDF-S, FB, SB, and HB. 
We did not stack the 
ISX and ISN sources in the 250~ks E-CDF-S because the total X-ray exposure is low
and no significant stacked signal can be achieved.
Sources located further than 6\arcmin\ from the 
average aim point are not included in the stacking due to
sensitivity degradation at large off-axis angles \citep[e.g.,][]{Lehmer2007,Xue2011}.
We visually examined the IR and optical images of the remaining
ISX and ISN sources, and excluded additional 10 sources 
(two ISX and eight ISN sources) whose 24~$\mu$m 
data appear to be affected by
source blending. The final ISX and ISN samples used in the stacking 
contain 23 and 58 sources, respectively.\footnote{Note that the entire sample
of ISX sources was selected in the E-CDF-S region which covers a total 
solid angle of $\approx1100$~arcmin$^2$, and thus the number of ISX 
sources within the inner 6\arcmin-radius area is only $\approx10\%$ of the total.} 
About 70\% of these sources 
have spec-$z$'s.
We also visually checked the X-ray images and verified that these
ISX and ISN sources are not close to any bright X-ray sources which
could affect the stacking results. 
Basic properties of the ISX and ISN sources used in the X-ray stacking 
analysis are listed in Table~1.

We followed a stacking procedure
similar to that discussed in \citet{Steffen2007}.  
For each source in each of the three standard bands, 
we calculated the total (source plus background) counts 
within a 3\arcsec-diameter circular aperture centered on its
optical position. 
This extraction radius was found to produce the best
signal-to-noise ratio (S/N) \citep[e.g.,][]{Worsley2006}
compared to other choices.
The corresponding 
background counts within this aperture were determined 
with a Monte Carlo approach which randomly (avoiding known \hbox{X-ray} sources) 
places 1000 
apertures within a 1\arcmin-radius circle of the source position
to measure the mean background \citep[e.g.,][]{Brandt2001}.
The total counts ($S$) and background counts ($B$) for the stacked
sample were derived by summing the counts for individual sources.
The net source counts are then given by $(S-B)$, and the S/N 
is $(S-B)/\sqrt{B}$; we note that the numbers of source and background counts are
large ($>100$), and thus the S/N can be calculated using Gaussian statistics.
The 3\arcsec-diameter aperture does not encircle
the full point-spread function (PSF).
Thus we determined an encircled-energy fraction (EEF) 
for each source in each band 
by interpolating an EEF table given the aperture radius, \hbox{off-axis} angle,
and photon energy. The EEF table was derived from 
the PSF images of the \hbox{main-catalog} sources in the 4~Ms CDF-S generated 
by {\sc acis extract} (AE; \citealt{Broos2010}) that uses the {\sc marx} ray-trace simulator;
PSF images at five different photon energies ($\approx0.3$--8.5 keV) for 
each \hbox{main-catalog} source in each
\chandra\ observation were provided
by AE (see \S3.2 of \citealt{Xue2011}). 
We created the EEF table by calculating the EEFs at different aperture radii, 
off-axis angles,
and photon energies, averaged over all the observations weighted by exposure time.
The EEFs calculated this way are the best representative of the 
real CDF-S data. 
The EEF in a 3\arcsec-diameter aperture
has a strong 
off-axis angle dependence, being $\approx95\%$
for on-axis sources, and 
$\approx50\%$ for sources at a 6\arcmin\ off-axis angle.
The aperture correction for the stacked counts was calculated using
the exposure-weighted average of the EEFs for all the sources.
The effective power-law photon indices and fluxes were calculated based on the band ratios
and aperture-corrected count rates 
using the CXC's Portable, Interactive, Multi-Mission Simulator
(PIMMS; see details in \S3.3.1 of \citealt{Luo2008}). 

The stacking results
are listed in Table~2, which show statistically significant differences between
the ISX and ISN samples. 
For the ISN sample, there is a strong detection ($10.6~\sigma$) in the SB, while
the stacked signal is weak ($1.6~\sigma$)\footnote{This is a marginal detection. The chance
of producing such a weak stacked signal by Poisson fluctuations is $\approx5\%$.
Treating this kind of weak signal as a detection does not affect our later analyses.} in the HB. The corresponding
band ratio is $0.29\pm0.19$ (1~$\sigma$ errors), which indicates an effective 
power-law photon index of $\Gamma=2.0\pm0.6$, 
consistent with X-ray emission from starburst galaxies \citep[e.g.,][]{Ptak1999}.
The ISX sample has a $\approx5~\sigma$ detection in the SB and
a $\approx4~\sigma$ detection
in the HB (corresponding to rest-frame $\approx3$--14 keV), 
with a band ratio of $1.5_{-0.5}^{+0.6}$, corresponding
to $\Gamma=0.6_{-0.4}^{+0.3}$. 
X-ray sources with very flat spectral slopes ($\Gamma<1$)
are almost exclusively identified as heavily obscured AGNs \citep[e.g.,][]{Bauer2004},
and thus
a significant contribution from obscured AGNs is
required to produce 
this kind of hard X-ray 
signal for the ISX sample.
We note that the ISN sample may still contain a fraction of 
low-luminosity AGNs with soft \hbox{X-ray} spectra comparable to those of star-forming galaxies
\citep[e.g.,][]{Gonz2006,Gonz2009}. These AGNs are generally
unobscured and are not of primary interest to this study. 
In Figure~3, we show the adaptively smoothed stacked images for the ISX and ISN
samples. It is clear that the ISX sample has 
a much harder X-ray signal than the ISN sample.
The difference between the ISX and ISN samples is also apparent via comparison
of their fluxes; the ISX SB flux is $\approx40\%$ smaller than the ISN SB flux, while the ISX HB flux 
is $\approx5$ times higher than the ISN HB flux.

\begin{figure}
\centerline{
\includegraphics[scale=0.5]{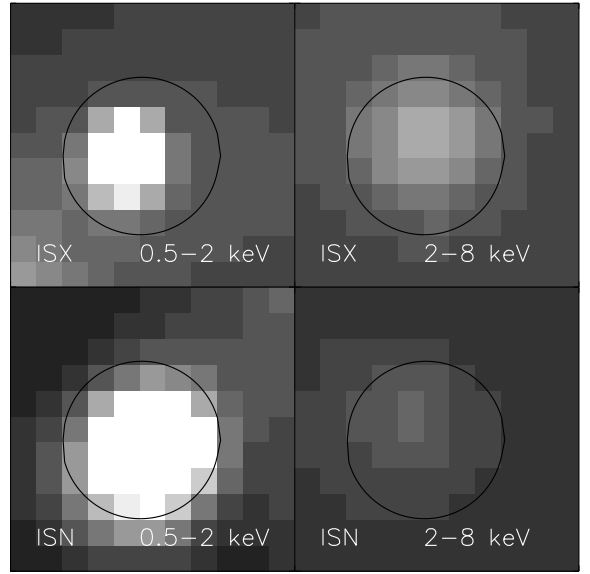}
}
\figcaption{
The adaptively smoothed stacked images for the ISX and ISN
samples. 
The left panels are for the SB, and the right panels are for the HB.
The images were adaptively smoothed with a minimal significance (S/N)
of 2.5~$\sigma$, and have been scaled
linearly with the same scaling. The black circles represent the 
3\arcsec-diameter aperture used to extract photometry. It is evident that
the ISX sample has a much harder stacked spectrum than the ISN sample.
}
\end{figure}

\subsection{Heavily Obscured AGN Fraction}

The stacked X-ray signal ($\Gamma\approx0.6$) for the ISX sample 
indicates the existence of embedded AGNs and is consistent with 
X-ray emission from heavily obscured or CT AGNs modeled with 
a reflection 
spectrum \citep[e.g.,][]{George1991,Maiolino1998} or a 
spectrum from a toroidal reprocessor \citep{Murphy2009a}.
However, we cannot derive the average AGN properties using the
stacked signal directly, as the ISX sources have a range of redshifts
and the sample is 
likely contaminated by star-forming galaxies
(see, e.g., Fig.~\ref{f01}). We thus performed Monte Carlo simulations to
assess the fraction of heavily obscured AGNs/CT AGNs 
in the sample, using a procedure 
refined from
that discussed in \citet{Fiore2009}. 
For a given AGN fraction, we ran 10\,000 simulations to
estimate the expected X-ray emission from both the star-forming
and AGN activities of the sample, 
and we then compared
the average simulated results to the observed stacked signal. 

\subsubsection{Simulation for the IR SFR Normal Sample}

We first used the ISN sample (58 star-forming galaxies) 
to test our simulation method.
For \hbox{star-forming} galaxies, a number of studies 
have found an approximately  
linear correlation between the SFR and
rest-frame 2--10 keV X-ray luminosity
$L_{\rm 2\textrm{--}10,gal}$ \citep[e.g.,][]{Bauer2002,Ranalli2003}.
However, later studies have pointed out that the X-ray--SFR correlation
is likely to have significant scatter due to the fact that SFR only
relates to the population of relatively young high-mass X-ray binaries (HMXBs),
and the older population of \hbox{low-mass} X-ray binaries (LMXBs)
likely relates to the galaxy stellar mass \citep[e.g.,][]{Colbert2004,Persic2007}.
Therefore, we adopted the most recent relation presented in \citet{Lehmer2010}, which
considers the correlation of $L_{\rm 2\textrm{--}10,gal}$ 
with both the SFR and stellar mass,
\begin{equation}
L_{\rm 2\textrm{--}10,gal}=\alpha M_*+\beta \textrm{SFR}~,
\label{sfrx}
\end{equation}
where $\alpha=2.33\times10^{29}$ erg~s$^{-1}$~$M_\sun^{-1}$ and 
$\beta=1.62\times10^{39}$~erg~s$^{-1}$~($M_\sun$~yr$^{-1}$)$^{-1}$;
here $\alpha$ has been increased by a factor of 2.6 to account for 
the average difference between stellar masses calculated
in this work and \citet{Lehmer2010}.
Ideally, the two SFRs for a given ISN
galaxy,
$\textrm{SFR}_{\rm IR+UV}$ and $\textrm{SFR}_{\rm UV,corr}$,
should be equal to each other. We consider that the Gaussian spread in 
$R_{\rm SFR}$ for ISN galaxies shown in 
Figure~2 is caused by uncertainties in
the two SFRs, and the intrinsic SFR is estimated as
$\log(\textrm{SFR})=[\log(\textrm{SFR}_{\rm IR+UV})+\log(\textrm{SFR}_{\rm UV,corr})]/2$.
The
corresponding 2--10 keV X-ray luminosity was then derived from Equation~\ref{sfrx},
also including a random Gaussian dispersion of 0.34 dex \citep{Lehmer2010}
to account for the scatter of that relation.

We assumed an X-ray power-law photon
index of $\Gamma=2.0$ (given the stacking results) 
and no intrinsic absorption for the star-forming galaxies, and then converted the 
predicted 2--10 keV X-ray luminosity to the observed SB and HB fluxes using the individual
redshifts of the sources.
As none of the ISN sources is individually X-ray detected,
we required that the simulated \hbox{X-ray} fluxes in the SB and HB of each source do not
exceed the sensitivity limits at the source position.
The
X-ray sensitivity limits are derived following \citet{Xue2011},
by calculating the minimum flux at each pixel (converted from the minimum number of counts)
required for detection under
the catalog source-detection criteria.
We added a small Gaussian dispersion of 0.1 dex to the sensitivity limits
to account for statistical fluctuations (estimated using the few CDF-S X-ray sources with
fluxes below the nominal sensitivity limits).
If the simulated X-ray fluxes are greater than the sensitivity limits,
we made another random generation of the source properties (in this case, only
Eq.~\ref{sfrx} and the sensitivity limits have random scatters).
The expected
SB and HB counts
were calculated given the exposure time, \hbox{flux-to-count-rate
conversion} (using PIMMS; see \citealt{Luo2008}),
and EEF of the source.
For each simulation, net counts of every
source were summed in the SB and HB as the output counts. We performed
10\,000 simulations, and the average numbers of output counts 
are $204.2\pm0.3$ for the SB and $57.4\pm0.3$ for the HB (the 
errors are the standard errors of the mean; the 1~$\sigma$ dispersions
for the average numbers of counts are both $\approx30$),
matching well with the observed 
stacked counts ($194.8\pm23.1$ and $52.4\pm33.0$).
Therefore we can successfully reproduce the X-ray emission
from star-forming activity
with this simulation approach.

\subsubsection{Simulations for the IR SFR Excess Sample}
We then performed
simulations for the 23 sources in the 
ISX sample adopting a similar approach, calculating
X-ray emission from both the star-forming
and AGN activities of the sample.
Assuming an AGN fraction in the range of $\approx10$--90\%, 
the X-ray emission from the star-forming 
galaxies in the sample was derived following the same procedure as above. 

For a source selected as an AGN, its X-ray emission and IR emission consist of both a 
star-formation component (host galaxy) 
and an AGN component. For the star-formation component,
we estimated its IR-based SFR to be 
$\textrm{SFR}_{\rm IR,sf}=k\times\textrm{SFR}_{\rm UV,corr}$, 
where the factor $k$ is randomly drawn from
a Gaussian function with $\sigma=0.33$ 
(blue histogram in Fig.~\ref{f01}) to account for
the scatter of the calculated SFR. 
The intrinsic SFR of the star-formation component was then estimated as
$\log(\textrm{SFR})=[\log(\textrm{SFR}_{\rm IR,sf})+\log(\textrm{SFR}_{\rm UV,corr})]/2$,
and the 
corresponding X-ray emission was calculated.
For the AGN component, the intrinsic
\hbox{rest-frame} 2--10 keV luminosity, 
$L_{\rm 2\textrm{--}10,AGN}$, can be estimated from the rest-frame 
6~$\mu$m luminosity of the AGN, $\nu L_{\nu,{\rm 6\mu m}}$ 
\citep[e.g.,][]{Lutz2004}. 
The relation in \citet{Lutz2004} was derived using a sample of local
AGNs with the 6 $\mu$m AGN continua decomposed from the IR spectra.
Here we used the luminosity-dependent version
of the relation which takes into account the possible luminosity dependence
\citep{Maiolino2007,Alexander2008},
\begin{equation}
L_{\rm 2\textrm{--}10,AGN}=10^{4.57}\times(\nu L_{\nu,{\rm 6\mu m}})^{0.88}~.
\label{irlxagn}
\end{equation}
A Gaussian dispersion of 0.4 dex was added to account for 
the scatter of the relation, estimated based on the 
range of intrinsic X-ray-to-mid-IR luminosity ratios for local AGNs \citep{Lutz2004}.
We note that this relation is consistent with the X-ray-to-mid-IR luminosity
ratios of a sample of X-ray selected obscured quasars
(Sturm et al. 2006), suggesting that it is applicable for deriving the intrinsic
luminosity of heavily obscured or CT AGNs.
To derive the rest-frame 6~$\mu$m monochromatic luminosity coming from the AGN, we first 
computed the AGN 24~$\mu$m flux by
subtracting the star-formation contribution from the observed 24~$\mu$m flux;
the star-formation contribution to the 24~$\mu$m flux was estimated using
the IR-based SFR ($\textrm{SFR}_{\rm IR,sf}$) above and the \citet{Chary2001} IR SED 
templates.
The residual 24~$\mu$m flux was then converted to the 
rest-frame 6~$\mu$m flux using the IR SED of the local CT AGN NGC~1068
(SED data from \citealt{Rigopoulou1999} and \citealt {Galliano2003}).
The observed 24~$\mu$m wavelength corresponds to rest-frame 
12--16~$\mu$m for our sources, and the conversion factor to the
rest-frame 6~$\mu$m flux is 
$\approx0.2$--0.3.

We adopted an intrinsic photon index of $\Gamma=1.8$ for the AGN and an
$N_{\rm H}$ value randomly drawn from the $N_{\rm H}$ distribution 
shown in Figure~7 of \citet{Gilli2007}, in which about 
half of the AGNs are CT. The adopted $N_{\rm H}$ distribution
has only a small impact on the final $N_{\rm H}$ distribution, as will be discussed 
later.
Given the intrinsic 2--10 keV luminosity, absorption, and photon index, we estimated
the observed absorbed SB and HB fluxes using the {\sc MYTorus} 
model \citep{Murphy2009a}\footnote{See \url{http://www.mytorus.com/} for details.}
implemented in XSPEC. The {\sc MYTorus} model is a recently developed 
spectral-fitting suite for modeling the transmitted and scattered X-ray spectra
from a toroidal-shaped reprocessor, and it is especially designed to 
treat the spectra of CT AGNs. Compared to the commonly used 
disk-reflection model for CT AGNs (e.g., the {\it pexrav} model in XSPEC; see 
\citealt{Magdziarz1995}), the {\sc MYTorus} model is more physically motivated
and takes into account absorption and scattering simultaneously.
We only considered the continuum output from {\sc MYTorus} for simplicity,
neglecting the fluorescent iron emission lines which contribute less than 8\% of the 
HB continuum flux. Two basic parameters describing 
the geometry of the model 
are the \hbox{half-opening} angle of the torus and the inclination angle (0\degr\ 
corresponds to a face-on observing angle). The \hbox{half-opening} angle was set at 37\degr,
corresponding to a scenario where the number of obscured AGNs is four times the number
of unobscured AGNs; such an obscured AGN fraction ($\approx80\%$) for moderate-luminosity
AGNs at $z\approx1$ 
has been reported in several studies \citep[e.g.,][]{Hasinger2008,Treister2008}.
We randomly chose an inclination angle between 37\degr\ and 90\degr\ (probability
weighted by solid angle) for each source. 
Other parameters, such as the relative cross-normalizations of
different components, were set at the default values (see \S8.2 of the {\sc MYTorus}
manual). 
The $N_{\rm H}$ value in the {\sc MYTorus} model has an upper limit of 
$10^{25}$~cm$^{-2}$. For column densities beyond this limit, 
the observed spectrum is highly geometry dependent, and the 
scattered flux is likely dominated by scattering in any optically thin
parts of the torus, even if the effective solid angle of the
optically thin gas is 
tiny (T. Yaqoob and K. D. Murphy 2011, private communication).
The above parameterization of 
the model cannot reproduce the exact environment of real obscured AGNs, 
which could be much more complicated \citep[e.g.,][]{Guainazzi2005,Comastri2010},
but we consider 
it to be the best available approximation of the average properties.
We then combined the emission emerging from this obscured 
AGN component with that from the star-formation component to derive the expected 
observed emission.

As for the simulations of star-forming galaxies, we required that
the observed SB and HB fluxes do not
exceed the sensitivity limits at the source position.
We would continue randomly
regenerating the source properties (e.g., $N_{\rm H}$ value)
until this requirement is satisfied.
This is a strong constraint on the simulated 
source properties, and it generally requires the
AGN to be heavily obscured or CT.
Finally, we converted the fluxes into counts in the extraction aperture,
and added the SB and HB counts from every source
as the output of a single simulation. We again performed 
10\,000 simulations, and the average output counts were 
compared to the stacking results in Table~2.

\subsubsection{Simulation Results}
The simulation results are shown in Figure~4, with the simulated SB and HB counts
at different assumed AGN fractions. 
As the AGN fraction increases,
the band ratio also increases and the simulated spectrum becomes harder.
We note that the simulated SB photons almost exclusively originate
from star-forming galaxies and AGN hosts (their star-formation components), 
while the HB photons are mainly from the AGNs; the increase of band ratio
with AGN fraction is largely due to the increase in the HB counts.
However, we did not include a soft-scattering component 
in the spectra of heavily obscured AGNs,
which 
has been observed to
provide significant SB emission in most of the local CT AGNs
\citep[e.g.,][]{Kinkhabwala2002,Guainazzi2009,Marinucci2011}.
The normalization of this
component to the intrinsic
power-law continuum is around 1\%
but has large object-to-object variations and 
uncertainties \citep[e.g.,][]{Comastri2007,Gilli2007,Ueda2007},
and the resulting contribution to the SB emission relative to the
host-galaxy contribution depends on both this normalization
and the intrinsic X-ray luminosity. 
Therefore 
the SB simulated counts should be considered lower limits, and 
we use only the HB results to determine the AGN fraction.
As shown in Figure~4,
an AGN fraction of $74\pm25\%$ can produce the observed HB counts 
(see Table~3).\footnote{We 
note that the $\approx25\%$ star-forming galaxies can still host
low-luminosity AGNs, as long as the AGN X-ray luminosities are not
comparable to their host galaxies; this kind of AGNs is not the focus of this 
study.}
The error on the AGN fraction
was determined given the error on the observed HB counts.
The $\approx75\%$ AGN fraction is in good agreement with the 
fraction of extra sources (73\%, shaded area) 
in the $R_{\rm SFR}$
distribution plot in Figure~2. 
For this best-fit AGN fraction, we also computed the stacked and simulated 
\hbox{X-ray} signals in
several sub-bands, and the comparisons are shown in Table~3.
The simulations reproduced well the stacked counts in general.
The cumulative $N_{\rm H}$ distribution for the 75\% of
AGNs in the 10\,000 simulations
is displayed as an inset of Figure~4, where $\approx95\%$ of the AGNs
are heavily obscured, and $\approx80\%$ are CT. The CT
AGN fraction in the entire ISX sample is thus $\approx60\%$.
The ISX AGNs have moderate X-ray luminosities.
The median intrinsic 2--10~keV luminosity derived from the simulations
is $L_{\rm 2\textrm{--}10,AGN}\approx2\times10^{42}$~\lum; this value would be
$\approx50\%$ higher if the intrinsic luminosity was calculated (Equation 6) using the 
observed IR flux (without subtracting the star-formation contribution).
The interquartile range of the X-ray luminosities
is (0.9--4)$\times10^{42}$~\lum\ [(1--5)$\times10^{42}$~\lum]
for the heavily obscured (CT) AGNs.\footnote{For
comparison to local studies, these
intrinsic luminosities are comparable to those of some well-known CT AGNs, such as
the Circinus galaxy and NGC 4945 (e.g., \citealt{Comastri2004} and references therein).}

We 
explored how the AGN fraction could change by varying some of the assumptions
in the simulations. 
First, we tried the commonly used disk-reflection model 
instead of the {\sc MYTorus} model
to compute obscured AGN emission.
An absorbed power-law ({\it zwabs*zpow} in XSPEC)
plus reflection ({\it pexrav} in XSPEC) model was used. The absorbed
power-law component typically dominates
in the Compton-thin ($N_{\rm H}\le1.5\times10^{24}$~cm$^{-2}$) regime, 
and the reflection component
dominates in the CT regime.
We note that the {\it pexrav} model has non-negligible uncertainties on some of its
parameters, such as the reflection scaling factor.
For simplicity, we adopted the same parameters used in \citet{Gilli2007} 
to model the XRB; we
assumed a cut-off energy of 200 keV, a reflection scaling factor of 0.37,
and an inclination angle of 60\degr\ for
the reflecting material.
The hard X-ray flux produced with this model is about three times higher than that
from the {\sc MYTorus} model \citep{Murphy2009a}, 
and thus the required AGN fraction in the ISX sample 
is smaller. The fraction drops from $74\pm25\%$ to $55\pm15\%$, still indicating a
substantial AGN contribution. Again, almost all ($\approx95\%$) of the AGNs are heavily obscured.
Simply increasing the
$L_{\rm 2\textrm{--}10,AGN}$-to-$\nu L_{\nu,{\rm 6\mu m}}$ ratio
in Equation~\ref{irlxagn} also reduces the AGN fraction.
For example, if we adopt the relation used in \citet{Fiore2009}, which
gives an X-ray luminosity that is $\approx0.2$ dex more luminous than ours,
the resulting AGN fraction is $\approx65\%$.
The initial $N_{\rm H}$ distribution input into the simulation will affect
the AGN fraction slightly. 
We adopted an alternative initial $N_{\rm H}$ distribution that has only $25\%$ 
CT 
AGNs, which is similar to
the $N_{\rm H}$ distribution for high
excitation-line galaxies in \citet{Tozzi2006}. The derived AGN fraction
is $60\pm15\%$ with $\approx90\%$ being CT. 
The CT AGN fraction 
in the entire ISX sample ($\approx55\%$) 
is thus comparable to that ($\approx60\%$) for the 
best-fit result above.
Even in an extreme case where the initial 
$N_{\rm H}$ distribution has only 10\% CT
AGNs, an AGN fraction
of $\approx60\%$ and a CT AGN fraction of $\approx55\%$
are still required to produce the stacked X-ray emission and 
satisfy the requirement that the simulated
sources cannot be individually detected in the SB or HB.
We note that the adopted photon index for star-forming 
galaxies ($\Gamma=2.0$)
has little effect on the estimated AGN fraction as these galaxies
produce mainly SB counts and
we do not
consider SB counts to be
constraining.

In general, the requirement that the simulated 
sources cannot be individually detected places a strong constraint on the final
results, and an AGN fraction of $\ga50\%$ (mostly heavily obscured) 
is always expected. We also caution that the simulation results 
were taken to be the 
average of 10\,000 tests, and thus they (as well as the following analyses
based on these results) 
only represent the most probable scenario of the ISX source properties
and might deviate from the real source properties.  
We consider that this is the best available 
approach to probe the nature of ISX sources   
given the X-ray and multiwavelength data available.

\begin{figure}
\centerline{
\includegraphics[scale=0.5]{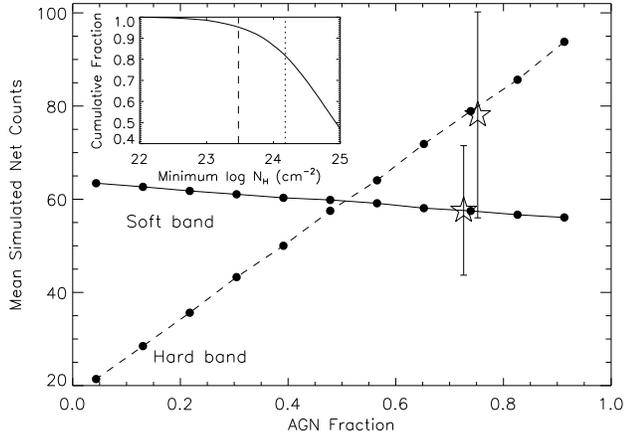}
}
\figcaption{
The average SB and HB counts of the ISX sample as a function of the 
AGN fraction, derived from Monte Carlo simulations. The statistical errors on the 
mean values are smaller than the symbol size. 
The SB simulated counts should be considered lower limits owing to
the highly uncertain soft scattering component not included in the simulations.
The two stars with error
bars 
indicate the observed stacked SB (lower star) and HB (upper star) 
counts, with the best-fit AGN fraction as the $x$-axis values. 
The HB (SB) 
data point is shifted rightward (leftward) by a small amount
for clarity.
An AGN fraction of 
$74\pm25\%$ can reproduce the observed signals. The inset panel shows the
cumulative $N_{\rm H}$ distribution at this AGN fraction, with the dashed and 
dotted lines indicating the criteria for heavily obscured and CT AGNs.
Approximately $\approx95\%$ of the AGNs in the ISX sample
are heavily obscured, and $\approx80\%$ are CT.
}
\end{figure}

\section{DISCUSSION}

\subsection{Space Density and Contribution to the 10--30 keV XRB} \label{density}
Given the simulation results for
the ISX sample, we can estimate the space density of heavily obscured AGNs 
at $z\approx0.5$--1 
and calculate their contribution to the XRB at high energies (10--30 keV).
There are $16\pm5$ ($13\pm4$) 
heavily obscured (CT) AGNs expected in the \hbox{X-ray} stacking ISX sample.
These objects
are located within the central 6\arcmin-radius region of the CDF-S; the
corresponding area is 0.029~deg$^{2}$ after excluding the regions masked by
X-ray sources. We further correct the number of sources by 10\% to
account for the fraction removed due to blended 24 $\mu$m photometry.
Therefore the sky density of heavily obscured (CT) AGNs in our ISX sample
is $\approx600$~deg$^{-2}$ ($\approx500$~deg$^{-2}$);\footnote{For simple comparison,
the sky density of $z=0.5$--1 X-ray AGNs from the original 5237 galaxies  
in the same region for the 
4~Ms CDF-S is 
$\approx800$~deg$^{-2}$.}
the corresponding space density is $(2.0\pm0.7)\times10^{-4}$~Mpc$^{-3}$
[$(1.6\pm0.5)\times10^{-4}$~Mpc$^{-3}$]
given the comoving volume between redshift 0.5 and 1. 

Note that there are several factors affecting the estimated space density
above. The expected number of AGNs in the ISX sample may be
up to $\approx30\%$ lower if different assumptions are made in the simulations (\S3.2),
and thus the space density will be reduced by the same fraction.
On the other hand,
there is probably a small fraction of sources excluded
from the parent sample with stellar mass $M_{*}\le5\times10^9~M_{\sun}$
(\S2.2) that also host heavily obscured AGNs. 
Also, the \hbox{X-ray} stacking ISX sample was selected in a relatively small area, and thus 
we have missed a few rare but intrinsically luminous
AGNs, which is also suggested by the moderate IR luminosities of 
all the ISX sources in Table 1.
Moreover, the X-ray spectra of heavily CT ($N_{\rm H}>10^{25}$~cm$^{-2}$) AGNs
cannot be modeled by {\sc MYTorus} and an upper limit of
$10^{25}$~cm$^{-2}$ was used for the column density; 
if these heavily CT AGNs are present in
the ISX sample, we could have overestimated their X-ray emission and thus
underestimated the AGN fraction and space density.
Most importantly,
we define the ISX sample conservatively ($R_{\rm SFR}>0.5$)
to avoid significant contamination from star-forming galaxies in the sample.
There will be additional heavily obscured AGNs among those sources that are not
in the ISX or ISN sample ($0.2<R_{\rm SFR}<0.5$;
see \S\ref{appsamples} below for X-ray stacking results);
the expected number is nearly comparable to that in the ISX sample
given the fraction of excess sources in the
distribution of $R_{\rm SFR}$ in Figure~2.
Taking into account the uncertainties and incompletenesses,
we expect that the true space density of
heavily obscured AGNs at $z\approx0.5$--1 is  
$\approx 1.5$--2 times that estimated above. In the following analyses, we still
use our conservative estimation as it is difficult to determine 
the exact correction factor.

We show in Figure~5 the space densities of
CT AGNs selected in this work and some previous studies (mainly IR based), 
as well as those from model predictions by \citet{Gilli2007}.
The CT AGN space density
for the \citet{Daddi2007} CDF-S sample with \hbox{$z\approx1.4$--2.5} and
$L_{\rm 2\textrm{--}10}\ga10^{43}$~\lum\ is estimated to be
$\approx2.6\times10^{-4}$~Mpc$^{-3}$; the actual value is expected to be lower
\citep[e.g.,][]{Donley2008,Murphy2009} and was recently revised 
downward to
$\approx2\times10^{-5}$~Mpc$^{-3}$ \citep{Alexander2011}.
At lower redshift, the space density for the \citet{Fiore2009}
COSMOS CT AGN sample with
$z\approx0.7$--1.2 is $(3.7\pm1.1)\times10^{-5}$~Mpc$^{-3}$.
We are probing a different population of CT AGNs from the \citet{Fiore2009}
COSMOS sample in terms of intrinsic X-ray luminosities.
The interquartile range of the intrinsic 2--10~keV luminosities
is (1--5)$\times10^{42}$~\lum\
for the CT AGNs in our sample,
while it is (3--10)$\times10^{43}$~\lum\ for the \citet{Fiore2009} sample.
We thus selected mainly moderate-luminosity and more typical CT AGNs;
this is largely attributed to the much higher sensitivities of the mid-IR and X-ray
observations in the CDF-S. On the other hand, we expect that we have missed some
rare, unrepresentative objects due to the smaller area of the CDF-S.
We note that the data points in various studies and model predictions have been 
derived with
different assumptions (see, e.g., \S3.2.2) and are thus not strictly comparable.

\begin{figure}
\centerline{
\includegraphics[scale=0.5]{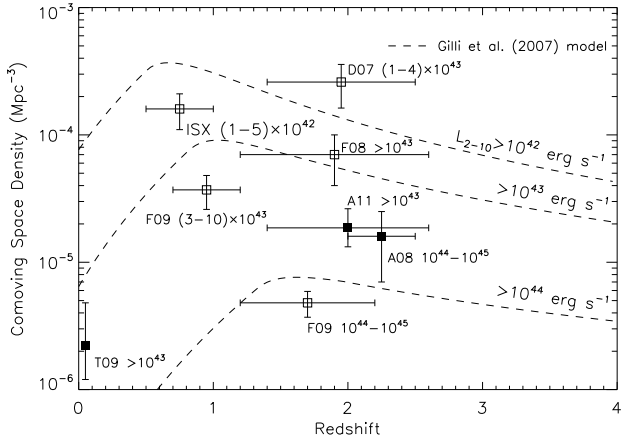}
}
\figcaption{
Space density of CT AGNs in the ISX sample. Also shown are the CT AGN 
space densities
in some previous IR-based studies: 
Daddi et al. (2007; D07), Alexander et al. (2008; A08),
Fiore et al. (2008; F08), Fiore et al. (2009; F09), 
and Alexander et al. (2011; A11). We also include the 
local CT AGN density from Treister et al. (2009; T09).  
The redshift error bars
indicate the redshift ranges of the samples, and the intrinsic 
2--10 keV luminosity range of
each sample is also shown. 
Note that the \citet{Alexander2008} and \citet{Alexander2011}
data points were derived from
spectroscopically (X-ray and/or IR) identified CT AGNs, 
the \citet{Treister2009}
data point was derived based on a sample of local CT AGNs,
and the other studies are based on \hbox{X-ray} stacking analyses of 
X-ray undetected candidates.
The
dashed curves are the predicted space densities of CT AGNs for 
different \hbox{X-ray} luminosity lower limits from the \citet{Gilli2007} population-synthesis
model. 
Note that the various data points and model predictions have been derived with
different underlying assumptions and are thus not strictly comparable.
\label{spacedensity}}
\end{figure}

We estimated the expected XRB flux in the 10--30 keV band provided by 
the heavily obscured AGNs in the ISX sample.
We used the {\sc MYTorus} model
to calculate the emergent flux from these AGNs, given their redshifts, 
assumed power-law photon index ($\Gamma=1.8$),
intrinsic X-ray luminosities, and column densities (the latter two quantities are
from the simulation results). On average, these AGNs produce a flux of 
$\approx0.6$~keV~cm$^{-2}$~s$^{-1}$~sr$^{-1}$ in the \hbox{10--30}~keV band, 
$\approx85\%$ of which is from CT AGNs. The total XRB flux in 
this band 
is about 44~keV~cm$^{-2}$~s$^{-1}$~sr$^{-1}$ (e.g., \citealt{Moretti2009} and references therein).
Therefore, this population of $z\approx0.5$--1
heavily obscured (CT) AGNs selected using the ISX method is expected to 
contribute $\approx1.4\%$ 
($\approx1.2\%$)
of the total XRB in the 10--30 keV band. 
Observations have not directly resolved the XRB at energies $>10$~keV;
however, 
theoretical predictions have been made by population-synthesis models about the 
contributions from unobscured, moderately obscured, and heavily obscured AGNs.
For example, \citet{Gilli2007} concluded that $\approx24\%$ of the XRB in the
10--30 keV band (the missing XRB) 
is produced by CT AGNs, while
\citet{Treister2009} considered the total contribution from CT AGNs to be
$\approx 9\%$.
The predicted contribution from the CT AGNs selected here ($\approx1.2\%$)
is thus
$\approx5\%$ \citep{Gilli2007} or $\approx15\%$ \citep{Treister2009}
of the missing XRB in the 10--30 keV band.
Given the expected properties (luminosity, redshift, column density) 
of the ISX sources, the \citet{Gilli2007} model predicts that $\approx7\%$ of 
the missing XRB comes from such objects,\footnote{\url{http://www.bo.astro.it/$\sim$gilli/xrb.html.}}
agreeing well with our results.
The remaining missing XRB is largely attributed to luminous
($L_{\rm 2\textrm{--}10}\approx10^{42}$--$10^{44}$~\lum) CT 
AGNs at $z\la1.5$ \citep{Gilli2007}; the \hbox{X-ray} stacking ISX sample selected here appears to 
have missed some of the high-luminosity
objects due to the limited volume surveyed.

\subsection{Resolved Fraction of the XRB by {\bf {\em Chandra}}}
Heavily obscured AGNs are expected also to contribute to the XRB 
at energies $<10$~keV, though with a smaller contributed fraction than that 
for the 10--30 keV band (see, e.g., Fig.~15b of \citealt{Gilli2007}). 
The XRB in the 1--8 keV band, unlike that in
the 10--30 keV band, has been largely resolved into discrete sources 
\citep[e.g.,][]{Bauer2004,Worsley2005,Hickox2006}.
With the deepest \chandra\ data available, the 4~Ms \hbox{CDF-S}, we 
expect to resolve the XRB further and improve our
understanding of 
the nature of the X-ray source populations 
at low flux levels, including the heavily obscured population.
We performed X-ray stacking analyses 
on the X-ray sources and optical galaxies in the CDF-S to explore the 
resolved XRB fraction and hidden AGN contribution to the 
unresolved fraction.

Measurements of the normalization of the XRB 
spectrum have non-negligible 
uncertainties and field-to-field variations; the combined uncertainty
on the normalization is $\approx10$--20\% \citep[e.g.,][]{Moretti2003,Hickox2006}. 
Here we adopted the normalization from \citet{Hickox2006}, 
which was derived from the \chandra\ Deep Fields data including the 
2~Ms \chandra\ Deep Field-North (CDF-N; \citealt{Alexander2003}) 
and 1~Ms CDF-S \citep{Giacconi2002}. 
The XRB has a power-law spectral slope with $\Gamma=1.4$ and a normalization of
10.9 photons~s$^{-1}$~keV$^{-1}$~sr$^{-1}$ at 1 keV.
X-ray stacking analyses were performed in
the following energy bands: 0.5--1, 1--2, 2--4, 4--6, and 6--8 keV.
For X-ray sources, we used the 740 sources in the 
4~Ms CDF-S main X-ray source catalog \citep{Xue2011}. 
The stacking procedure was 
the same as that described in \S3.1; to maximize the
S/N ratio, a 3\arcsec-diameter circular aperture was used and the stacking was performed 
for the 389 sources within the inner 6\arcmin-radius area.  
To account properly for bright X-ray sources that have a rare occurrence 
in the narrow CDF-S region, 
we adopted the bright-end correction in \citet{Hickox2006}:
sources brighter than $5\times10^{-15}$~\flux\ in the SB or
$1.4\times10^{-14}$~\flux\ in the HB are removed from the stacking, and
the background intensity produced by such bright sources was calculated using 
the number counts of X-ray sources \citep{Hickox2006}.
Four sources in the SB and two in the HB are removed this way; the final stacking
samples include 385 sources in the 0.5--1 and 1--2 keV bands, and 387 sources in
the three $>2$ keV bands.
For the optical galaxies, we chose $z$-band sources 
in the GOODS-S {\it HST} version r2.0z catalog \citep{Giavalisco2004}\footnote{See
http://archive.stsci.edu/pub/hlsp/goods/catalog\_r2/.} with
a 5 $\sigma$ limiting AB magnitude of
$28.2$. 
We
stacked the X-ray undetected optical
galaxies in the central 6\arcmin-radius region.
Galaxies within twice the
90\% encircled-energy
aperture radius of any known X-ray source are also removed from the stacking 
to avoid X-ray source contamination.
There are 18\,272 optical sources included in the stacking. Note that
this galaxy sample contains all of the 23 sources in the X-ray stacking
ISX sample.

\begin{figure}
\centerline{
\includegraphics[scale=0.5]{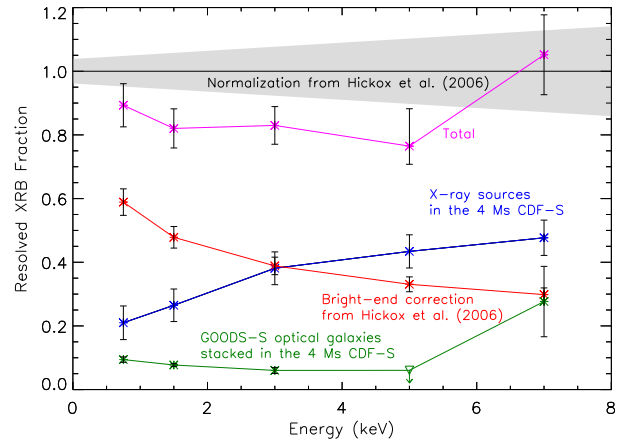}
}
\figcaption{Resolved fractions of the XRB in five energy bands between 0.5 and 8 keV. 
The total XRB intensities are  
from \citet{Hickox2006} with uncertainties indicated by the gray region. The stacked
contributions from X-ray sources in the 4~Ms CDF-S, \hbox{bright-end} correction,
GOODS-S optical galaxies, and the sum of the above are shown as blue, red, dark green,
and magenta data points, respectively. Note that the stacked signal for the 
optical galaxies in the 4--6 keV band did not yield a significant detection and thus
a 3~$\sigma$ upper limit on the resolved fraction was calculated 
(triangle with a downward arrow). The upper limit was used to derive the upper error
when calculating the total resolved fraction in this band.
For the other data points, 1~$\sigma$ errors are shown.
The contribution from the 
ISX objects 
is $<1.2\%$ in the 6--8 keV band and is not shown.
[{\it See the electronic
edition of the Journal for a color version of this figure.}]
\label{xrbfrac}}
\end{figure}

The stacking results are shown in Figure~\ref{xrbfrac}. 
The 1~$\sigma$
errors on the stacked fluxes were
calculated following \cite{Hickox2006} including measurement errors and
a 3\% \chandra\ flux-calibration error. For the stacking of the X-ray sources,
there is an additional Poisson error
due to the limited number of sources below the bright-end flux cut \citep{Hickox2006}.
Taking into account both the X-ray source contribution and bright-end
correction, the resolved XRB fractions are $\approx75\%$--80\% in all the
energy bands, indicating that the average photon index of the
X-ray sources is $\Gamma\approx1.4$. This is consistent with the fact that
the majority of the X-ray sources in the CDF-S are obscured AGNs \citep[e.g.,][]{Luo2010,Xue2011}.
For the X-ray undetected optical galaxies,
significant detections are found
in all the bands except the 4--6 keV band, where there is only a 1~$\sigma$ signal
(corresponding to a $\approx20\%$ chance that the signal was created by Poisson noise). 
We thus calculated 3~$\sigma$ upper limits on the stacked counts and resolved XRB fraction in
this band. In the 6--8 keV band, the optical galaxies produced a 2.5~$\sigma$ stacked 
signal ($\approx1\%$ chance of being generated by Poisson noise), responsible for 
$28\pm11\%$ of the XRB.\footnote{As a check of the stacking method, we 
stacked 18\,272 random positions (excluding X-ray sources) 
in the central 6\arcmin-radius region, and the stacked signals in all the bands 
are consistent with zero ($<1~\sigma$ significance).}
With all of the X-ray sources and galaxies considered,
it appears that the XRB in the 6--8 keV band can be fully explained, 
though the uncertainties in the normalization and stacked signals are large
(the slightly higher stacked flux than the XRB flux in this band could also
be caused by cosmic variance in the CDF-S).
At 1--6 keV, X-ray sources and galaxies can account for $\approx80$--$90\%$
of the XRB.
The remaining unresolved fraction
is probably due to cosmic variance in
the narrow CDF-S region \citep[e.g.,][]{Bauer2004,Luo2008}. It 
could also be partially 
contributed by extended X-ray sources 
(e.g., galaxy groups or clusters) in the CDF-S.

The 2.5~$\sigma$ stacked signal from galaxies in the 6--8 keV band
($28\pm11\%$ of the XRB; we refer to this
as the galaxy CT XRB),
combined with the weak signal in the
4--6 keV band
($<6\%$ of the XRB), suggest 
an underlying population of heavily obscured AGNs among the X-ray undetected
galaxies; emission in the 4--6 keV band is more heavily absorbed than that in
the 6--8 keV band.
As a subsample of the optical galaxies,
the 23 ISX sources produce a
stacked signal that is $0.2\pm0.1\%$ of the total XRB flux in the 4--6 keV band;
the stacked signal in the 6--8 keV band is weak (1.4~$\sigma$), and we 
estimated a 3~$\sigma$ upper limit that is $1.2\%$ of the total XRB.
Therefore, in the 6--8 keV band, the ISX sources contribute $<5\%$ of the
galaxy CT XRB.
It is expected that there are
some additional heavily obscured AGNs not selected with the ISX method 
among the optical
galaxies (see \S4.1 and \S4.3). 
The number density of such objects
is of about the same order of magnitude as the ISX sources.

Based on the \hbox{population-synthesis} model in \citet{Gilli2007}, we derived that
heavily obscured AGNs below the 4~Ms \hbox{CDF-S} sensitivity limit\footnote{We 
adopted here the 
median 2--8 keV sensitivity limit within the inner 6\arcmin-radius 
region of the 4~Ms CDF-S \citep{Xue2011}.} are
responsible for $\approx8\%$ of the XRB in the 6--8 keV band (we refer to this
model prediction as the model CT XRB).
This model CT XRB is only $\approx30\%$ of the
galaxy CT XRB,
due to possible cosmic variance in the \hbox{CDF-S} and uncertainties in the 
model assumptions (e.g., assumptions about the spectral shape,
CT AGN number density, and obscured fraction; 
see \S9 of \citealt{Gilli2007} for discussion).
We also note that the \citet{Gilli2007} model does not include 
low-luminosity ($L_{\rm 2\textrm{--}10}\la10^{42}$~\lum)
AGNs. IR studies suggest that the AGN fraction  
in low-mass galaxies 
may be significantly higher than previously  
reported using optical spectroscopy \citep[e.g.,][]{Goulding2009}.
The obscured AGN fraction also increases as luminosity 
decreases \citep[e.g.,][]{Hasinger2008}.
It is thus probable that low-luminosity heavily obscured AGNs
have a significant stacked
contribution to the galaxy CT XRB.
These AGNs are faint in the X-ray band and challenging to identify
even in deep \chandra\ observations \citep[e.g.,][]{Goulding2010}.
Their IR luminosities are more likely to be dominated by host-galaxy emission
and thus are difficult to detect with the ISX method (or any IR-based method).

\subsection{Additional Samples and Subsamples} \label{appsamples}

We extracted three additional X-ray stacking samples with $0.2<R_{\rm SFR}<0.5$ 
to test whether there are AGNs among these sources.
We further break the ISX sample into two subsamples
to explore how the stacking results 
depend on the $R_{\rm SFR}$ threshold value.
The threshold cuts in $R_{\rm SFR}$ 
for these samples (samples A1--A5) are listed in Table~2, and
they were chosen so that each sample has a similar number ($\approx10$) 
of sources.  
We performed
X-ray stacking analyses on the two subsamples and three additional samples 
following the same procedure as that for the ISX and ISN samples.
The results are presented in Table~2. Due to limited sample sizes,
these samples have less-significant detections in the HB than the ISX sample. 
However,
the band ratios and effective photon indices for samples \hbox{A1--A4} all suggest
heavily obscured AGN contributions to the stacked X-ray signals.
A plot of the effective photon index as 
a function of the average $R_{\rm SFR}$ is shown in Figure~7;
in general, the
larger the $R_{\rm SFR}$ threshold value (more IR excess),
the harder the stacked signal.
Similar behavior 
has also been observed by \citet{Daddi2007}, and it is likely due to
less contamination from star-forming galaxies at larger threshold values
(see Fig.~2). Given the stacking results for these additional samples, 
it is not likely that the hard
X-ray stacked signal of the ISX sample was produced by coincidence.

\begin{figure}
\centerline{
\includegraphics[scale=0.5]{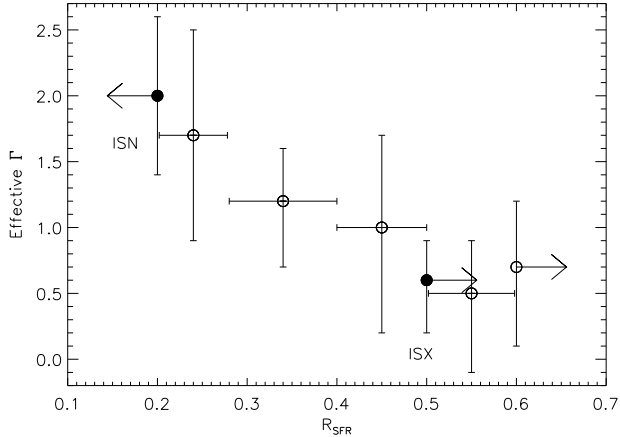}
}
\figcaption{
The effective photon index of the stacked X-ray signal vs. 
the average $R_{\rm SFR}$ of the sample. 
Besides the ISX and ISN samples (filled circles), several additional 
samples (open circles) are presented with different threshold 
cuts in $R_{\rm SFR}$ (see Table~2). The error bars or arrows in
the $x$-axis direction indicate the intervals of the $R_{\rm SFR}$ cuts.
In general,
as the average $R_{\rm SFR}$ increases, the stacked X-ray spectrum gets flatter, 
suggesting more significant contributions
from heavily obscured AGNs.
}
\end{figure}

\subsection{Observational Prospects for Distant Heavily Obscured AGNs}

A potentially straightforward way to detect distant heavily obscured or CT AGNs
is via hard \hbox{X-ray} observations at $\approx10$--100~keV. 
Several future hard X-ray
missions, such as {\it NuSTAR} (planned launch year 2012; \citealt{Harrison2010}) 
and {\it ASTRO-H} (planned launch year 2014; \citealt{Takahashi2010}), have as one science
goal to detect hard X-ray emission from distant heavily obscured AGNs. For the three typical
hard X-ray bands of {\it NuSTAR}, \hbox{6--10~keV}, 10--30 keV, and 30--60~keV, the
expected sensitivity limits are $\approx2\times10^{-15}$, $2\times10^{-14}$,
and $6\times10^{-14}$~\flux\ for a 1~Ms exposure,\footnote{See \url{http://www.nustar.caltech.edu/}.} respectively.
These flux limits are about two orders of magnitude
more sensitive than those of previous missions.
The 
\hbox{{\it ASTRO-H}} sensitivities are expected to be comparable to those for {\it NuSTAR}. 
The angular resolutions (half-power diameters) of {\it NuSTAR} and \hbox{{\it ASTRO-H}} are expected
to be $\approx50\arcsec$ and $\approx1.7\arcmin$, respectively. 
The {\it NuSTAR} positional accuracy is expected to be $\approx1.5\arcsec$ for 
strong sources.
To assess if the AGNs in our ISX samples at $z\approx0.5$--1 will
be detectable, we utilized the properties for sources in the X-ray stacking ISX sample from
the best-fit simulations (for the 74\% AGN fraction) and the {\sc MYTorus} 
model to calculate
their expected fluxes in the {\it NuSTAR} bands.
Only a tiny fraction ($\approx0.4\%$) of the simulated sources in our 10\,000 simulations
have fluxes above the
{\it NuSTAR} sensitivity limit in its most sensitive band (\hbox{10--30~keV});
this corresponds to no detected sources expected
within the $16\pm5$ heavily obscured AGNs
in the \hbox{X-ray} stacking ISX sample (or one detectable source among
all the 242 ISX objects).
The median simulated flux in the \hbox{10--30}~keV band ($\approx3\times10^{-16}$~\flux)
is about two orders of magnitude below the sensitivity limit.
This result is a natural consequence of the
requirement that the sources are not individually detected in the 2--8 keV band of the 4~Ms
\chandra\ exposure; the extraordinarily high sensitivity of the 4 Ms CDF-S places a tight
constraint on the intrinsic luminosities of the ISX sources and prevents them from being
detected by {\it NuSTAR}. Therefore, it is not likely that {\it NuSTAR} or 
{\it ASTRO-H} will detect
any of the ISX sources presented here.
However, they will probably detect some of the X-ray selected CT AGN candidates
\citep[e.g.,][]{Tozzi2006,Comastri2011} in hard X-rays; such detections will be
useful for a clear
determination of the intrinsic spectral shape and power of these sources.

One other approach to identify heavily obscured or CT AGNs is via
X-ray spectroscopy at relatively low energies ($<10$~keV) 
complemented by multiwavelength data \citep[e.g.,][]{Polletta2006,Tozzi2006,Alexander2008,Alexander2011,Comastri2011}.
The X-ray emission of these objects is characterized by a flat continuum
and often a strong
rest-frame 6.4~keV iron K$\alpha$ fluorescent line \citep[e.g.,][]{Della2008,Murphy2009a}.
For the ISX sources presented
here, spectroscopic
analyses are probably not feasible due to the small number of counts expected. However,
it is worth performing a
further study of the \hbox{X-ray} selected CT AGN candidates in \citet{Tozzi2006}, which
were previously studied using only the 1~Ms CDF-S data. With the current 4~Ms CDF-S data and the
3~Ms \hbox{\it XMM-Newton} observations on the CDF-S, we will be able to constrain better
their nature \citep[e.g.,][]{Comastri2011}. 
In the case of the CDF-S receiving further \chandra\ exposure (e.g., 10~Ms total), 
some ($\approx15\%$) of the 
heavily obscured AGNs in the ISX sample could be detected in the HB, given the 
simulated properties and expected 10~Ms sensitivity. 

Given the above, the majority of the heavily obscured AGNs in the ISX 
sample will remain undetected
in the X-ray, and the few percent of the XRB at $\approx10$--100~keV produced by these
AGNs cannot be directly resolved in the near
future. We note that X-ray absorption variability appears common among local Seyfert
2 galaxies \citep[e.g.,][]{Risaliti2002}, and a few
sources have even exhibited CT to Compton-thin transitions \citep[e.g.,][]{Matt2003,Risaliti2005}. 
Therefore, a fraction 
of the ISX AGNs may be detectable in the future 
if they become less obscured due to absorption variability. 
Also, optical/mid-IR spectroscopy has been shown to be a
promising technique for identifying distant heavily obscured AGNs,
particularly when combined with sensitive X-ray constraints
\citep[e.g.,][]{Alexander2008,Gilli2010}.
Future optical/IR instruments such as the {\it James Webb Space Telescope} 
({\it JWST}),  Extremely Large Telescopes 
(ELTs), and Atacama Large Millimeter Array (ALMA) will provide great opportunities for detecting the remaining sources.

\section{CONCLUSIONS AND FUTURE WORK}

We have identified a population of 
heavily obscured/CT AGNs at $z\approx0.5$--1 in the CDF-S
and E-CDF-S utilizing the ISX selection method.
The key points from this work are listed below:

\begin{enumerate}
\item 
We have improved the ISX method from
\citet{Daddi2007} by deriving the dust extinction
via SED fitting and studying sources at lower redshifts with
no bias in their IR-luminosity calculation.
The ISX method can be well applied to the X-ray undetected
galaxies in the CDF-S region at $z\approx0.5$--1, given the superb
multiwavelength data available. The parent sample of 1313 galaxies
were defined from an initial sample of 5237 optically and 24~$\mu$m selected
sources in the E-CDF-S, with the requirements of selected redshift interval, 
excellent multiwavelength coverage, high stellar mass, and no 
X-ray detection (\S2). 

\item 
We have identified 242 ISX sources in the CDF-S and E-CDF-S at $z\approx0.5$--1;
these sources tend to be hosted by evolved galaxies with high stellar mass
and little dust (\S2.2).
An X-ray
stacking analysis of 23 of the objects in the central CDF-S region
resulted in a very hard X-ray signal with an effective photon
index of $\Gamma=0.6_{-0.4}^{+0.3}$, indicating
a significant contribution from
obscured AGNs (\S3.1).

\item
We have performed Monte Carlo simulations to estimate the AGN fraction in the ISX sample
and assess the intrinsic properties of the obscured AGNs.
We modeled the
observed X-ray flux considering
both the star-formation
and AGN contributions, and we utilized the {\sc MYTorus} model
to treat the spectra of heavily obscured AGNs.
The requirement that the sources are not individually detected in the 4 Ms CDF-S
sets strong constraints on their intrinsic properties.
We infer that $74\pm25\%$ of the ISX sources are obscured
AGNs, within which $\approx95\%$ are heavily obscured and $\approx80\%$ are CT (\S3.2).

\item
The heavily obscured (CT) AGNs discovered in our ISX sample
have moderate intrinsic X-ray luminosities;
the interquartile range of the intrinsic 2--10~keV luminosities
is \hbox{(0.9--4)$\times10^{42}$}~\lum\ [(1--5)$\times10^{42}$~\lum].
The space density of the heavily obscured (CT) AGNs
is $(2.0\pm0.7)\times10^{-4}$~Mpc$^{-3}$ [$(1.6\pm0.5)\times10^{-4}$~Mpc$^{-3}$].
These heavily obscured (CT) AGNs are expected to contribute 
$\approx 1.4\%$ ($\approx1.2\%$)
of the total XRB flux in the 10--30 keV band.
In the 6--8 keV band, the 23 ISX sources provide $<1.2\%$ of the XRB flux.
The X-ray undetected optical
galaxies 
in the CDF-S produce a 2.5~$\sigma$ stacked signal in the 
6--8 keV band, accounting for $28\pm11\%$ of the XRB flux, which
is about the entire unresolved XRB fraction.
The space density of the ISX AGNs and their contribution to the XRB
could be increased by a factor of $\approx1.5$--2 due to our conservative 
ISX definition (\S4).

\item
These heavily obscured or CT AGNs will probably not be detected by hard X-ray
observatories under development such as
{\it NuSTAR} or {\it ASTRO-H} due to their moderate intrinsic X-ray luminosities
and significant obscuration. Most of the hard X-ray sources that will be detected by 
these facilities are likely already detected in the 4 Ms CDF-S given the 
extremely high sensitivity of \chandra\ to point sources (\S4.4).

\end{enumerate}

The ISX selection method presented in this paper can be applied 
to other survey fields with good optical-to-IR coverage, and it can be expanded
to higher redshifts if the IR luminosity can be estimated reliably, e.g.,
combining MIPS 24~$\mu$m data 
with data from deep {\it Herschel} surveys at 100 and 160~$\mu$m 
(e.g., \citealt{Shao2010}).  
We will explore these 
possibilities in future work.

~\\
We acknowledge financial
support from CXC grant SP1-12007A (BL, WNB, YQX),
CXC grant G09-0134A (BL, WNB, YQX),
NASA ADP grant NNX10AC99G (WNB), 
the Science and Technology Facilities Council (DMA), 
and ASI/INAF grant I/009/10/0 (AC, CV, RG).
We are grateful to T. Yaqoob and K. D. Murphy for providing support 
for the {\sc MYTorus} model and making available the data for our desired 
half-opening angle. We thank A. T. Steffen for helpful discussions.
We also thank the referee for carefully
reviewing the manuscript and providing helpful comments that
improved this work.

%\bibliographystyle{apj} 
%\bibliography{refs} 

\begin{deluxetable}{lcccccccccc}
\tabletypesize{\footnotesize}
\tablewidth{0pt}
\tablecaption{List of ISX and ISN Sources in the X-ray Stacking Analysis}

\tablehead{
\colhead{RA} &
\colhead{Dec}&
\colhead{$z$}&
\colhead{$z$ Lower} &
\colhead{$z$ Upper}  &
\colhead{$A_V$}&
\colhead{$f_{24}$}&
\colhead{$\log L_{\rm IR}$} &
\colhead{$\textrm{SFR}_{\rm IR+UV}$} &
\colhead{$\textrm{SFR}_{\rm UV,corr}$} &
\colhead{$R_{\rm SFR}$} \\
\colhead{(1)} &
\colhead{(2)} &
\colhead{(3)} &
\colhead{(4)} &
\colhead{(5)} &
\colhead{(6)} &
\colhead{(7)} &
\colhead{(8)} &
\colhead{(9)} &
\colhead{(10)} &
\colhead{(11)} 
}
\startdata
\multicolumn{11}{c}{ISX Sample} \\
& & & & & & & \\
\hline
  03 32 49.61 &$-$27 49 00.1  & 0.98& 0.97 &1.02& 0.2 &  58.7 & 10.74 &  6.2 &  1.1 & 0.76 \\
  03 32 45.58 &$-$27 49 36.4  & 0.680 & $-$1.00& $-$1.00& 0.9 & 472.3 & 11.46 & 29.5 &  5.0 & 0.77 \\
\hline
& & & & & & & \\
\multicolumn{11}{c}{ISN Sample} \\
& & & & & & & \\
\hline
  03 32 50.38 &$-$27 47 07.1  & 0.537 & $-$1.00& $-$1.00&1.9 & 166.5 & 10.86 &  7.7 & 14.1&$-0.26 $ \\
  03 32 48.58 &$-$27 45 04.9  & 0.89& 0.89&0.92& 1.2 & 121.8 & 11.03 & 12.2 & 12.2&$ 0.00 $ \\
\enddata
%\footnotesize
\tablecomments{
Table~1 is presented in its entirety in the electronic edition. An
abbreviated version of the table is shown here for guidance as
to its form and content.
The full table contains 11 columns of information
for the 23 ISX sources and 58 ISN sources used in the X-ray stacking analysis.
Cols. (1) and (2): The J2000 right ascension and declination of the
ISX or ISN source. 
Cols. (3)--(5): The spec-$z$ or photo-$z$ of the source.
Spec-$z$'s are denoted by having three decimal
places, while photo-$z$'s with their 1~$\sigma$ confidence intervals 
(lower and upper bounds) were derived using ZEBRA.
Col. (6): The $V$-band dust extinction derived from ZEBRA SED fitting.
Col. (7): The MIPS 24~$\mu$m flux, in unites of $\mu$Jy.
Col. (8): The logarithmic 
IR (8--1000~$\mu$m integrated) luminosity estimated based on the observed
24~$\mu$m flux, in units of solar luminosity.
Cols. (9) and (10): The IR-based and UV-based SFRs, in units of $M_{\sun}$~yr$^{-1}$.
Col. (11): The logarithmic ratio of the IR-based and UV-based SFRs.
}
\end{deluxetable}

\begin{turnpage}

\begin{deluxetable}{lcccccccccccccc}
\tabletypesize{\scriptsize}
\tablewidth{0pt}
\tablecaption{Stacked X-ray Properties}

\tablehead{
\colhead{} &
\colhead{}&
\colhead{}&
\colhead{}&
\multicolumn{3}{c}{Net Source Counts} &
\multicolumn{3}{c}{Signal-to-Noise Ratio} & 
\colhead{} &
\colhead{} &
\multicolumn{2}{c}{Flux}&
\colhead{}\\
\colhead{} &
\colhead{}&
\colhead{}&
\colhead{}&
\multicolumn{3}{c}{\rule{2in}{0.01in}} &
\multicolumn{3}{c}{\rule{1.2in}{0.01in}} &
\colhead{Band}&
\colhead{Effective}&
\multicolumn{2}{c}{\rule{0.7in}{0.01in}}  &
\colhead{} \\
\colhead{Sample} &
\colhead{$N_{\rm gal}$} &
\colhead{$z_{\rm mean}$} &
\colhead{$t_{\rm exp}$ (Ms)}&
\colhead{FB} &
\colhead{SB} &
\colhead{HB} &
\colhead{FB} &
\colhead{SB} &
\colhead{HB} &
\colhead{Ratio} &
\colhead{$\Gamma$} &
\colhead{SB} &
\colhead{HB} &
\colhead{$L_{\rm X}$} \\
\colhead{(1)} &
\colhead{(2)} &
\colhead{(3)} &
\colhead{(4)} &
\colhead{(5)} &
\colhead{(6)} &
\colhead{(7)} &
\colhead{(8)} &
\colhead{(9)} &
\colhead{(10)} &
\colhead{(11)} &
\colhead{(12)} &
\colhead{(13)} &
\colhead{(14)} &
\colhead{(15)}
}

\startdata
& & & & & & & & & &\\
%\vspace{0.1 in}
ISX ($R_{\rm SFR}>0.5$)& 23& 0.69& 76.7& $138.8\pm26.1$& $57.6\pm13.9$& $78.1\pm22.1$& 6.0& 4.9 &3.9 &$
1.48_{-0.51}^{+0.63}$&$0.6_{-0.4}^{+0.3}$&4.9& 36.6&8.3\\
ISN ($R_{\rm SFR}<0.2$)& 58& 0.77& 193.0 & $256.8\pm40.3$& $194.8\pm23.1$& $52.4\pm33.0$& 6.9& 10.6 &1.6&
$0.29\pm0.19$&$2.0\pm0.6$&7.7&7.3 &4.2\\
%\hspace{0.1 in}
& & & & & & & & & &\\

\hline
%\vspace{0.1 in}
& & & & & & & & & &\\
A1 ($R_{\rm SFR}>0.6$) & 14&0.68&46.9&$71.5\pm20.0$ &$31.4\pm10.6$ &$39.0\pm17.0$&4.0
&3.5&2.6& $1.35_{-0.68}^{+0.90}$ &$0.7_{-0.6}^{+0.5}$&4.2&28.4&6.3 \\
A2 ($0.5<R_{\rm SFR}<0.6$) & 9&0.71&29.8&$67.3\pm16.7$ &$26.1\pm9.0$ &$39.1\pm14.1$&4.5
&3.5&3.2& $1.64_{-0.73}^{+1.04}$ &$0.5_{-0.6}^{+0.4}$&5.9&50.7&12.2 \\
A3 ($0.4<R_{\rm SFR}<0.5$) & 10&0.76&33.3&$45.8\pm16.8$ &$25.1\pm9.1$ &$20.7\pm14.1$&3.1
&3.2&1.8& $0.91_{-0.67}^{+0.81}$ &$1.0_{-0.8}^{+0.7}$&5.6&22.7&7.1 \\
A4 ($0.28<R_{\rm SFR}<0.4$) & 11&0.69&37.7&$81.7\pm18.5$ &$47.7\pm10.6$ &$33.0\pm15.2$&5.1
&5.9&2.4& $0.76_{-0.37}^{+0.41}$ &$1.2_{-0.5}^{+0.4}$&8.8&28.1&7.4 \\
A5 ($0.2<R_{\rm SFR}<0.28$) & 13&0.74&43.7&$69.4\pm19.4$ &$49.2\pm11.2$ &$19.7\pm15.9$&4.0
&5.6&1.3& $0.44_{-0.36}^{+0.37}$ &$1.7\pm0.8$&8.1&12.7&5.0 \\
\enddata
\footnotesize
\tablecomments{Col. (1): The X-ray stacking sample. The threshold
cut in $R_{\rm SFR}$ is indicated.
Col. (2): Number of sources used in the stacking.
Col. (3): Mean redshift of the stacked sample.
Col. (4): Total FB exposure time.
Cols. (5)--(7): Stacked net source counts in the FB, SB, and HB, with
1~$\sigma$ Gaussian statistical errors.
Cols. (8)--(10): Stacked signal-to-noise ratios in the FB, SB, and HB.
Note that there are a few marginal detections ($<2$~$\sigma$) in the HB.
Treating this kind of weak signals as detections does not affect our analyses
in the paper.
Col. (11): Stacked band ratio for the stacked sample, 
defined as the ratio of count rates
between the HB and SB. The 1~$\sigma$ errors were calculated following the 
``numerical method'' described in \S1.7.3 of \citet{Lyons1991}.
Col. (12): Effective photon index with 1~$\sigma$ errors 
for the stacked sample.
Cols. (13)--(14) SB and HB fluxes for the stacked sample, in units of
$10^{-18}$~erg cm$^{-2}$ s$^{-1}$.
Col. (15): Rest-frame 0.5--8 keV X-ray luminosity calculated from the observed-frame
0.5--8 keV flux 
for the stacked sample, 
in units of $10^{40}$~\lum.
The mean redshift, observed flux, and effective power-law photon index were used in the calculation, and no assumption was made about the intrinsic absorption.}
\end{deluxetable}

\end{turnpage}

\begin{deluxetable}{lcc}
\tablewidth{0pt}
\tablecaption{Comparison of Stacked and Simulated 
Counts for the ISX Sample}
\tablehead{
\colhead{Band (keV)} &
\colhead{Stacked Counts}&
\colhead{Simulated Counts}
}

\startdata
0.5--2.0 & $57.6\pm13.9$ & $57.5\pm0.2$ \\
2.0--8.0  & $78.1\pm22.1$ & $78.9\pm0.3$ \\
0.5--1.0 &$22.6\pm8.6$ & $19.9\pm0.2$ \\
1.0--2.0 & $33.2\pm11.0$ & $37.6\pm0.3$  \\
2.0--4.0 & $31.6\pm13.3$ &$36.8\pm0.4$ \\
4.0--8.0 & $44.7\pm17.6$ &$42.1\pm0.4$ \\
\enddata
\tablecomments{The uncertainties for the stacked counts
are 1~$\sigma$ Gaussian statistical errors.
The simulated counts are the average values over the 10\,000 simulations;
see \S3.2 for details.
}
\end{deluxetable}

\end{document}